\documentclass[review]{elsarticle}

\usepackage{graphicx}
\usepackage{comment}
\usepackage{booktabs}
\usepackage{setspace}
\usepackage{ragged2e}
\usepackage{caption}
\usepackage{rotating}
\usepackage{wrapfig}
\usepackage{xurl}
\usepackage{xcolor,colortbl}
\usepackage{float}
\usepackage{lineno,hyperref}
\modulolinenumbers[5]

\journal{Journal of \LaTeX\ Templates}

\begin{document}

\begin{frontmatter}

\title{To remove or not remove Mobile Apps? \\A data-driven predictive model approach}

\author{Fadi Mohsen $^*$  \href{mailto:f.f.m.mohsen@rug.nl}{f.f.m.mohsen@rug.nl}\\, Dimka Karastoyanova
 \href{mailto:d.karastoyanova@rug.nl}{d.karastoyanova@rug.nl }\\
and George Azzopardi
 \href{mailto:G.Azzopardi@rug.nl }{g.azzopardi@rug.nl }}
\address{Information Systems Group,\\
Bernoulli Institute for Mathematics, Computer Science and Artificial Intelligence, \\University of Groningen, 9712 CP Groningen, The Netherlands}



%\ead{f.f.m.mohsen@rug.nl}


\begin{abstract}
Mobile app stores are the key distributors of mobile applications. They regularly apply vetting processes to the deployed apps. Yet, some of these vetting processes might be inadequate or applied late. The late removal of applications might have unpleasant consequences for developers and users alike. Thus, in this work we propose a data-driven predictive approach that determines whether the respective app will be removed or accepted. It also indicates the features' relevance that help the stakeholders in the interpretation. In turn, our approach can support developers in improving their apps and users in downloading the ones that are less likely to be removed. We focus on the Google App store and we compile a new data set of 870,515 applications, 56$\%$ of which have actually been removed from the market. Our proposed approach is a bootstrap aggregating of multiple XGBoost machine learning classifiers. We propose two models: user-centered using 47 features, and developer-centered using 37 features, the ones only available before deployment. We achieve the following Areas Under the ROC Curves (AUCs) on the test set: user-centered = 0.792, developer-centered = 0.762. 
\end{abstract}

\begin{keyword}
Third-party apps, Mobile apps, App stores, Actions, Broadcast receivers, Privacy, Permissions, XGBoost, Predictive Analysis, Android.
\end{keyword}

\end{frontmatter}

\section{Introduction}
\label{sec:introduction}
\begin{figure}
    \centering
    %\captionsetup{justification=centering,margin=2cm}
    \includegraphics[width=1\textwidth]{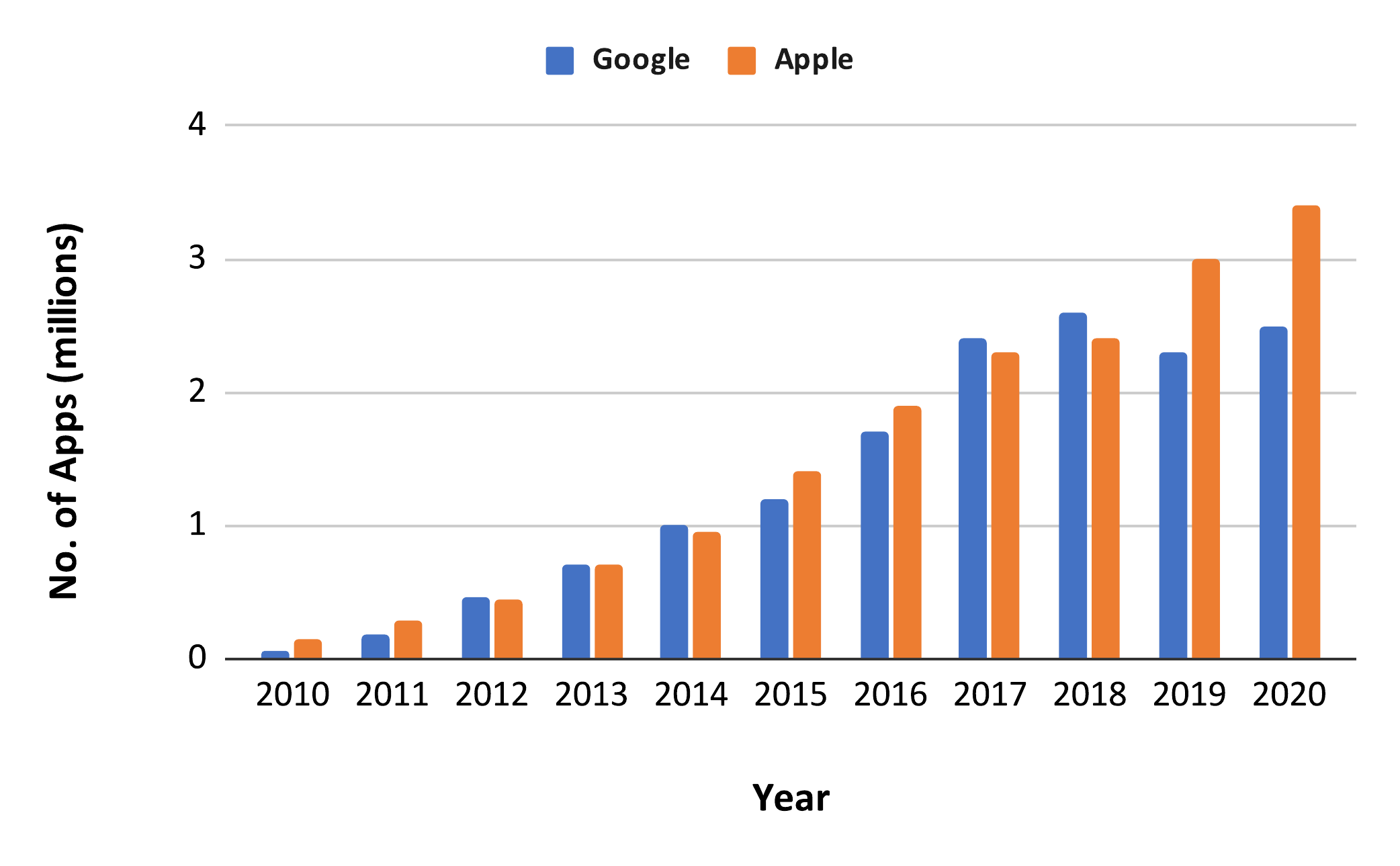}
    \caption{The number of apps in the Google Play and Apple stores between 2010 and 2020 \cite{CurryMarkets21}.}
    \label{fig:twodmarkets}
\end{figure}

%how can we differentiate between malicious and non-adhering apps, when both of them get removed
The mobile-app industry has grown tremendously in the last decade and is expected to keep rising. For example, Figure~\ref{fig:twodmarkets} shows the number of applications in two popular app stores, Google Play and Apple. The number of applications between 2010 and 2020 has enormously increased, from thousands into millions. This growth has also been accompanied with an increased number of malware and vulnerable applications \cite{8587187, 8714014,7173047,8718344, bba5d9341b2a427192a2347fc775610f}. In response to these threats, researchers have proposed numerous defense solutions to protect the privacy of end users \cite{8985418,8423084} and the security of their devices \cite{9000578}. Additionally, the mobile app stores have also implemented quality and security check measures to combat the different threats which resulted in removing a lot of applications from both markets between 2017 and 2019 \cite{Wottrich2018ThePT}. %Figure~\ref{fig:twodmarkets} shows that the Apple store had 0.1 million growth between 2017 and 2018 in comparison to 0.4 million growth the year before. Similarly, the growth of the Google Play store in the same year was also noticeably low. Moreover, between 2018 and 2019, it was the first time the store sees a decline in the number of apps, dropping from 2.6 to 2.3 million \cite{CurryMarkets21}.
Legitimate mobile app stores have long been compared against each other based on numerous factors such as submission process, cost, and the amount of guidance that is given to developers. The Google Play store, for instance,  was criticised at first for not rigorously vetting apps before approving and making them available to users \cite{iOSPlay2020}. 
%For some App Stores, instead of aiming at the privacy of mobile users and security of their devices, their vetting is mainly concerned with whether apps meet certain guidelines \cite{PlayStoreAgrement19} pertaining the user interface, privacy statements, description of the app, among others. The duration of the vetting process ranges from a couple of hours to few days. For instance, Apple is known to be stricter in the approval process than Google, an iOS app review would normally take up to 2 days in comparison to three hours for an Android application.
%https://www.businessofapps.com/data/app-stores/

As such, Google has been introducing a number of solutions to monitor its app store, which resulted in removing a large number of applications \cite{8595206}. For example, they introduced Google Play protect \cite{PlayProtect2020}, which is responsible for the rejections, removals, and suspensions of violating and suspicious third-party applications. The system issues also warnings and sometimes delivers push notifications to the developers of removed or suspended apps. 

Both Google Play Developer Distribution Agreement \cite{PlayStoreAgrement21} and Google Play Program Policies \cite{DeveloperPolicyCenter21} contain extensive details and instructions regarding what should and should not be included in mobile applications. The precise requirements are, however, still unclear and prone to misinterpretation, needless to say that some developers ignore these guidelines. Moreover, the Google Play store gives less guidance when an app is rejected in comparison to the iOS store \cite{iOSPlay2020}.
%As a result, their apps get flagged and removed from the store for violating the store policies. Though, in such situations not every developer understands the true reason why their application has been removed.
%In order to ensure the quality of
%newly added applications, Google created some policies and guidelines to regulate and
%qualify these apps.
%When deleting unsuitable applications, Google relies on %Google Play protect, built-in
%malware protection for Android. The system is responsible for rejections, removals,
%suspensions, warnings and sometimes delivery of a push notification to the developers
%of a removed or suspended app.
%The large-scale study by Wang et al. \cite{8595206} revealed that almost half of the studied apps have been removed or replaced from the Google Play store during a two-year period from 2015 to 2017. 

Removing violating apps have various negative consequences on both benign developers and mobile users. This is especially the case if the applications were removed  from the Google Play after they have been admitted and made available to users. 
%Both Google Play Developer Distribution Agreement \cite{PlayStoreAgrement19} and Google Play Program Policies \cite{DeveloperPolicyCenter20}
%contain extensive details and instructions regarding what should and should not
%be included in an app before it is allowed into Google Play
%Store. However, the precise requirements are still unclear and prone to
%misinterpretation. 
When an app is removed, the notifications sent by Google Play protect \cite{PlayProtect2020} are very generic and do not give developers any particular directions as to how to fix their apps. As a result, not just their apps get flagged and removed from the store, their accounts might also get suspended. On the other hand, it is also inconvenient for users when some of the apps they have been using get removed from the store. % draw some stats from the dataset that supports this claim, e.g. the average number of users who have downloaded apps that got removed
%we don't actually make any distinction between malicious and bad (does not follow the guidelines).
%Not only developers, but also users can benefit from such model/system to compare apps in terms of quality

Researchers have long studied the factors that influence the trustworthiness of mobile apps in online stores. Because of that, numerous frameworks have been proposed to assess their trustworthiness \cite{6523820}, risk \cite{10.5555/2534766.2534812}, quality and suspicious behavior \cite{7283021, DBLP:conf/icse/SlavinWHHKBBN16}. Determining the removability of mobile apps from the App stores is a challenging problem because there are numerous potential reasons as to why mobile applications get removed from these stores. In addition, it is very challenging if not impossible to enumerate all of these causes. Moreover, some of these reasons are not easy to pinpoint automatically or identify statistically \cite{8595206}. Thus, researchers in their efforts to tackle this problem have considered fewer reasons and relied on manual analysis of the removed apps.
%, and made certain assumptions. For example, 
For example, Wang et al. %conducted an empirical study on a large number of mobile apps collected from the Google Play store to understand why some of them are being removed. They
relied heavily on the manual analysis of the removed Android apps \cite{8595206}. Their machine learning classifier was only focused on COPPA-violated apps. Similarly, Seneviratne et al. \cite{10.1145/2736277.2741084} also relied on manual analysis of the collected app samples, focused only on detecting spam apps, and assumed that top apps with respect to the number of downloads, number of user reviews, and rating, are quite likely to be non-spam. 

%In their study, they manually analyzed  According to that study, more than half of the apps (out of 1.5 million) were removed from the store within two years. Among the removed apps were 500 popular ones with more than 1 million downloads each. The removed apps were labelled into different categories: malicious, privacy-risk, fake, spamming, and privacy-violating.

%Thus, in this work, we aim to propose two generic models for predicting the removability of apps in stores. 

The aim of our work is, thus, to develop two data-driven predictive models that can determine whether a given app will be removed or maintained by the Google Play store \textit{before} its deployment and \textit{after} it has been deployed. The predictive models are based on a machine learning algorithm called Extreme Gradient Boosting, or XGBoost \cite{chenatal16}. It leverages a mix of contextual and technical app's features such as the privacy policy link,  the genre, the requested permissions or privileges, and broadcast listeners. The models are meant to support developers, users, and app stores. We expect that they will help developers determine whether their apps are likely to be removed or not; hence, giving them an opportunity to review and fix their apps before submitting them to the store. In addition, they can assist users in choosing applications that are less likely to be removed. Lastly, Google Play store may consult with these models to identify violating applications early on before admitting them into the store or afterwards. It is worth noting that we do not apply any manual analysis on the collected samples. In addition, our models are generic meaning that they do not include any domain-specific considerations; they are purely data-driven.
%to worry about violating apps, instead focuses on removing truly malicious and inappropriate applications from the stock.

Our contribution in this work is threefold: First, we generated a very large data set of mobile applications from the Google Play store that includes the meta data, the Android Package (APK) files\footnote{APK is the package file format used by the Android operating system.}, and most importantly their standing in the store for over a year. The generated large data set is used to evaluate our approach, and is made publicly available \cite{H0YJFT_2022}.
%\footnote{The data set can be found here: https://cutt.ly/DQ0ib8W}. 
Second, we propose two predictive models -- developer-centered and user-centered -- that can indicate whether or not an app will be kept in the Google app store. 
%To the best of our knowledge we are the first to propose a solution to this problem. 
We believe that our data set and our encouraging results can be considered as a benchmark for further investigations.
Third, we present different usage scenarios of the two models, in which they can be integrated into a service or an app. In this work, we follow the CRISP-DM methodology to address our research question, which is spanned across three sections: Methodology, Experiments and Results, and Discussion.

This paper is organized as follows: Section~\ref{sec:bg} provides general information about the topic,  Section~\ref{sec:relatedwork} introduces relevant works in the literature; Section~\ref{sec:methodology} presents our data set and proposed method; Section~\ref{sec:experiments} describes the experiments that we conducted; Section~\ref{sec:discussion} discusses the results; and Section~\ref{sec:conclusion} presents our conclusions.
\section{Background}
\label{sec:bg}
We lay out the necessary background information regarding Android mobile applications, namely their distribution format, configuration file, permissions, and broadcast receivers.
\subsection{Distribution}
\label{subsec:dist}
Android applications are distributed via official and non-official app markets in the Android Package file format (\texttt{.apk}). Official markets such as the Google Play store and Samsung Galaxy store apply a number of quality and safety checks on the admitted apps. Nonetheless, malicious and low-quality applications are frequently being published into these stores and downloaded by a large number of users. 
Additionally, Android applications can be downloaded from anywhere online. Because of that, the default security settings of Android is set to disallow installation of such applications, from unknown external sources.
Since the official Google Play store is not available in some countries, alternative stores came out to fill that gap, such as the Xiaomi App Store, the 360 Mobile Assistant and the Huawei App store.
\subsection{Android Manifest File}
\label{subsec:dist}
Any APK file usually contains among other things a configuration file called \texttt{AndroidManifest}. The manifest file of an app
contains its configurations such as permissions, package name, broadcast receivers, and main activity. Extracting this file out of the APK file is easier and more accurate than retrieving the original source code \cite{androidobf}. In addition, accessing the manifest files of installed applications on mobile phones is also possible and accurate \cite{PackageManager2021}. This is why researchers have long relied on it for building security and privacy solutions as we will detail in Section~\ref{sec:relatedwork}.
\subsection{Android Permissions}
\label{subsec:perm}
The majority of third-party Android applications require some level of access to the device resources such as \texttt{SMS, Contacts} and \texttt{Camera}. The Android OS uses a permission system~\cite{PermissionsOnAndroid} to control the level of access each installed application has. The developers of these apps are, therefore, obligated to enlist all the permission requests in their apps' \emph{AndroidManfiest.xml} files.
\subsection{Broadcast Receivers}
\label{subsec:bcr}
In Android OS, a broadcast receiver of system actions allows apps to listen to events originating from the system. Examples include receiving SMS, call or voicemail, or when the WIFI is connected. As such, broadcast receivers might cause some security and privacy concerns to the end user. In Section~\ref{sec:relatedwork} we discuss previous studies that employed broadcast receivers in assigning privacy scores to apps.
\section{Related Work}
\label{sec:relatedwork}
%There is a diverse number of existing works concerning the quality of mobile apps. According to our specific application domain and to provide a clear structure, we categorize the existing works into two main categories:
%criticize existing research based on when they are applied 
%elaborate more to make it 2 columns
Our research aims at helping users, developers, and app stores' maintainers. As such, part of the discussion of the literature will be referring to these beneficiaries. Furthermore, the bulk of research that have been conducted in this area is focused on one or more type of bad applications. 

Wang et al. \cite{8595206} identified five categories, namely malicious, privacy-risk, fake, spamming, and privacy-violating. The models that we propose in this work are meant to be more generic, in that they will forecast whether an app will be maintained or removed, without specifying the reason of removal.  
%policy-violating, intrusive, and malicious applications. 
Lin et al. \cite{10.1145/3442381.3449990} conducted a similar study but on iOS app store. Their work was focused on understanding the reasons why apps are being removed from the app store. Their data set was based on collecting daily snapshot of the whole iOS app market for a year and a half. They also built app removal prediction model based on a number of features that are extracted from
the app metadata. Their model does not consider the apps that are new to the market because it simply relies on features that do not get populated immediately, e.g., app comments. Consequently, the model cannot be used by developers to predict the future of their apps before they upload them to the market.

The work of Wang et al. \cite{8595206} is the closest to our work, in which they wanted to understand why some apps are being removed from the store. They implemented an empirical study on a large number of mobile apps collected from the Google Play store. However, the status of each app in their data set was only checked once, a year and a half later. In our work, the status was checked on three different occasions: the first check was done after 5 months, the second was done after 7 months, and the last was done after 1 year.
%as shown in Figure~\ref{fig:dscollection}.
Their manual analysis of the collected apps identified a set of 791,138 removed apps. %In their work, the apps samples were analyzed manually. 
%In addition, their proposed machine learning classifier focuses only on COPPA-violated apps. 
They then ran an existing machine learning classifier \cite{10.1145/2873587.2873597} on this set to detect COPPA-violated apps, more specifically, apps targeting kids. Out of the 791,138 removed apps, the classifier has identified a total number of 23,319 apps targeting kids.
It is important to note that this work aimed at encouraging researchers to build symptom-based predictor or even a machine learning-based predictor for flagging the to-be-removed apps before they are really removed. Thus, our work is an improvement of their work since it employs more sophisticated techniques, relies on more features, and one of our models, the developer-centered model, is designed to be effective even before the app gets submitted to the store. 

Seneviratne et al. \cite{10.1145/2736277.2741084} proposed an Adaptive Boost classifier to detect spam apps based solely on their metadata that are available at the time of publication. Their work inspired us to use two of their features, namely \emph{IsSpamming} and \emph{DeveloperCategory}. Though, their classifier considers only spam apps, our two models do not distinguish between the different categories of bad applications. For us, an app is either removed or not. Additionally, their work relied on manual analysis of the collected app samples and assumed that the considered top apps with respect to the number of downloads, number of user reviews, and rating, are quite likely to be non-spam.

Researchers have long studied the factors that influence the trustworthiness of mobile applications in online stores. For example, Kuehnhausen and Victor ~\cite{6523820} proposed a trustworthy model for mobile applications based on various factors, namely ratings, permissions, reviews and the relationships between applications. However, the number of features that were used in building the model is relatively small. Additionally, the ratings and the reviews features could be empty for some apps, especially, if these apps have not been long in the market or are not popular. Finally, the data set that was used to evaluate their proposed model is small and does not sufficiently represent the entire market because the focus was on popular apps. In our work, however, we use a much bigger and more representative data set, more features, and investigate two approaches; one that relies on features available before deployment and another that uses features that become available after deployment.

Natural Language Processing (NLP) techniques were investigated by Pandita et al. \cite{10.5555/2534766.2534812} on the description of an app and compared it with the permissions that the app had requested. Their aim was to examine whether the description of an application provides any indication for why the application needs a permission. In our view, we believe that in order to justify the use of a permission by an app, more features are needed besides the description such as the genre and system actions. Pratim Sarma et al. \cite{10.1145/2295136.2295141} on the other hand, used the genre to inform users whether the risks of installing an app is in accordance with its advertised benefit. In our work, the description, genre, permissions, system actions and more features are incorporated in the models.

A framework to help the user deciding whether a given application found in some app stores is trustworthy or not was introduced by Habib et al. \cite{8455897}. It considers the publicly available information of an app such as user ratings and reviews, and also indicators regarding the security and as provided by state-of-the-art static analysis tools. As we explained earlier, the ratings and the reviews' features could be empty, but they did not consider such cases like we do.

The permissions an app requests, the category of the app, and the permissions that are requested by other apps in the same category to infer a privacy score that would be used to help users with their installation decisions were leveraged by Sarma et. al \cite{Sarma:2012:APP:2295136.2295141}. 
The work of Mohsen et al. \cite{8456016} did also devise a new privacy score for mobile applications, which is calculated based on the permissions they possess, the system actions they have registered to listen to and the users' privacy preferences. Both scores \cite{Sarma:2012:APP:2295136.2295141, 8456016} could have been improved if more of the application's meta data was utilized.

A large-scale longitudinal study on 5 million app records collected from three different snapshots of the Google Play store was conducted by Wang et al. \cite{10.1145/3308558.3313611}. Their study revealed a number of serious issues in the mobile app ecosystems. For example, the study shows that despite Google's effort to remove bad apps from the store, the number of developers who do not comply with the guidelines has been nonetheless increasing. In our view, their results highlight the need to have a solution like the one we propose in this paper.  
%Seneviratne et al.  

%For the purpose of helping app developers check their privacy policies against their internal implementation for consistency, Wang et al. \cite{DBLP:conf/icse/SlavinWHHKBBN16} proposed a framework that links policy phrases to sensitive API methods to detect anomalies. In our work we did not crawl the policy documents, instead we crawled the URLs that point at them. We then treated each URL as binary; either present (1) or absent (0). We speculate that analyzing privacy policies might improve further the accuracy of our predictor, something that we will investigate in future work. 

%Sanz et al. \cite{6181075} "we propose a new method for categorising Android applications through machine-learning techniques. "
% \begin{figure*}[t]
% \centering
% \captionsetup{justification=centering,margin=1cm}
%     \fbox{\includegraphics[trim=30 15 18 15 , clip,width=\textwidth]{Figures/Dataset Collection (3).png}}
%     \caption{A high-level overview of the data collection workflow.}
%     \label{fig:dscollection}
% \end{figure*}

Researchers have also used some of the contextual and technical features that we used in our research but for detecting malicious Android applications. For example, Peiravian and Zhu \cite{6735264} proposed to combine permissions and API (Application Program Interface) calls to detect malicious Android apps. Wu et al. \cite{6298136} considered static information including permissions, deployment of components, intent messages passing and API calls for characterizing the Android applications' behavior. Sanz et al. \cite{doi:10.1080/01969722.2013.803889}, Sato et al. \cite{Sato2013DetectingAM}, Feldman et al. \cite{7035780}, and Li et al. \cite{7790261} extracted and used several features from the Android manifest of the applications to build machine learning classifiers for the detection of Android malware. In our work, we decided not to use any source code related features, because obtaining such features in real life is complicated, especially if the applications are paid and/or obfuscated.

Gómez et al. \cite{7283021} analyzed the permissions and the user reviews of mobile applications to detect defective applications. The proposed system is aimed at helping app store maintainers predict apps with bugs before they harm the reputation of the app store as a whole.

Our work differs from the existing literature in three main aspects. First, we acquire and curate a very rich data set, that is large in volume and in independent variables. Moreover, the value of the dependent variable (removed or not removed) has been checked three times over a period of 1 year. Second, we propose two types of models that we refer to as developer-centered and user-centered, which can be applied before and after deployment, respectively. Third, our approach is completely data-driven. This means that the interaction between the given independent variables is learned from the training data rather than being imposed or manipulated by domain-specific knowledge.
% \begin{figure*}[t]
% \centering
% \captionsetup{justification=centering,margin=2cm}
%     \includegraphics[width=0.80\textwidth]{Figures/Dataset Collection (3).png}
%     \caption{A high-level overview of the data collection workflow.}
%     \label{fig:dscollection}
% \end{figure*}

\section{Methodology}
\label{sec:methodology}
In this section, we describe the data set that we collect and curate followed by feature engineering and the predictive models.
%the steps we followed to build the proposed data-driven predictive model. We first begin with the collection and curation of the data set before moving into defining the features that we use in our approach. 
%\begin{figure}
%\centering

%\includegraphics[width=8cm,height=6cm]{Figures/%boxplot-reviewsavg.png}
%    \caption{The distribution of the reviews average that was given by users to all applications.}
%    \label{fig:reviewsavg}
%\end{figure}

%\subsection{Overview}
 %Additionally, in Figure~\ref{fig:modeloverview}, we show a high-level overview of the steps we took to create and evaluate our approach. %In the subsequent sections, we will be referring to these two figures. 

%\label{sec:overview}
\subsection{Data set}
\label{sec:dataset}
%In this Section, we describe the data collection process, which is shown in Figure []. We first created a Google Play crawler to index the apps that are listed in the Google Play store \footnote{https://play.google.com/store}, and downloaded the apps' contextual information such as app description, genre, ratings, and privacy URLs. Using this crawler, We were able to collect a dataset of over 2 million Android applications. The dataset contained paid and free applications. In the next stage and using another crawler, we were able to download the APK files of nearly half of these applications, only the free ones. The dataset were collected between April 2017 and November 2018. We then checked the status of these applications on three different occasions; December 2018, February 2019, and May-June 2019. The purpose was to check which of the apps were still in the market and which got removed. 
%\subsubsection{Preprocessing}
%\label{sec:preprocessing}
%\subsubsection{Collecting the data set}
\begin{figure*}[t]
\centering
\captionsetup{justification=centering,margin=1cm}
    \includegraphics[trim=30 15 18 15 , clip,width=\textwidth]{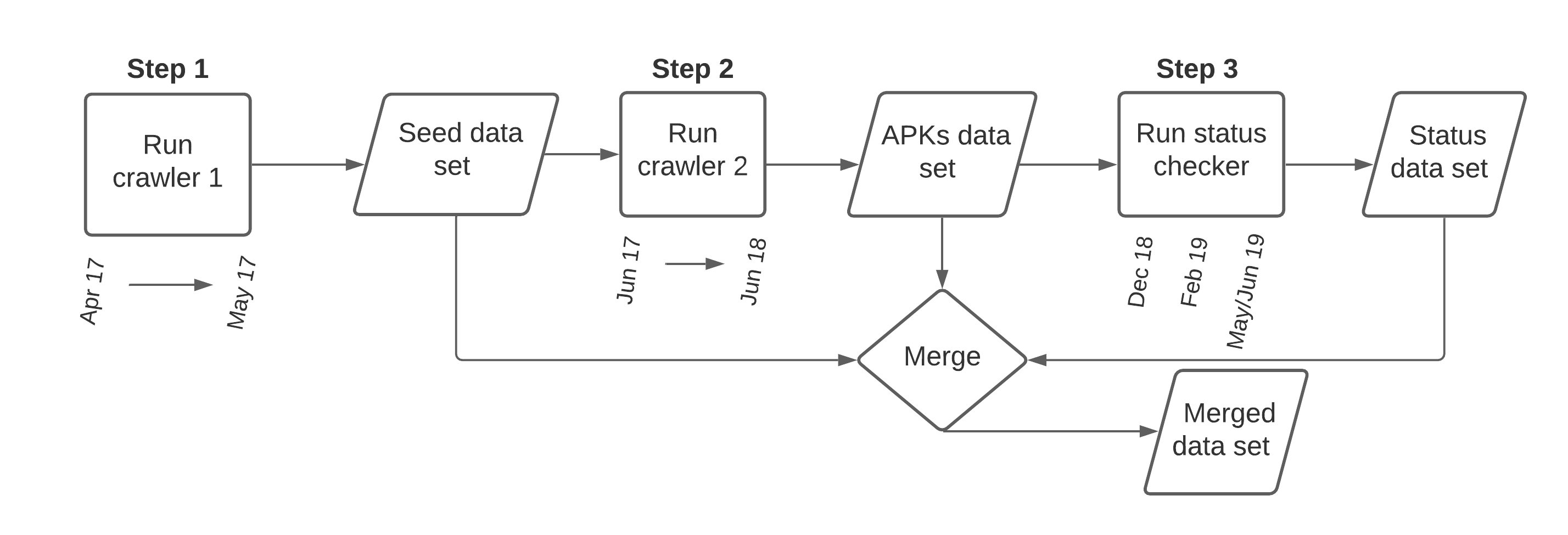}
    \caption{A high-level overview of the data collection workflow.}
    \label{fig:dscollection}
\end{figure*}

In Fig.~\ref{fig:dscollection} we show our methodology for collecting and curating the data set that we used for training and evaluating our models. We collected the data over a course of 26 months, between April 2017 and June 2019. In the first one-month long phase, we crawled the Google Play store main pages which resulted in 2,021,159 applications including all the information that we could find on these pages except for the permissions and users' reviews. We call this collection the \emph{seed} data set, the output of \emph{Step 1} in Fig.~\ref{fig:dscollection}. Then, in the second phase, which lasted a year, we downloaded the APK files for slightly more than half of the apps that are in the \textit{seed} data set. We call this collection the \emph{APKs} data set, which contains 1,164,216 apk files, \emph{Step 2} in Fig.~\ref{fig:dscollection}. 
%From the APK files, in particular the manifest configuration files, we extracted new features concerning the permissions and system actions. 
The last phase was geared towards collecting the ground truth labels (i.e. the values of the dependent variable) for all apps in the \textit{APKs} data set. It involved checking the status of the apps in the app store to see whether they are still in there or got removed. This phase was executed on three different occasions. We call this list the \emph{status} data set, which contains 1,090,484 apps, \emph{Step 3} in Fig.~\ref{fig:dscollection}.
%; the first check was done during the month of December 2018 for all apps in the apks data set, a second check was conducted also for all apps in the apks data set during the month of February 2018, and finally, the status of all apps in the seed data set was verified one last time during the months of May and June 2019. We are planning on making these data sets available for researchers to conduct further research activities.

% \begin{figure}
% \centering
%     %\includegraphics[width=0.48\textwidth]{Figures/barchart-lowestandvers.png}
%      \includegraphics[width=0.48\textwidth]{Figures/andvers.pdf}
%     \caption{A histogram that shows the distribution of the lowest Android version that was specified in all apps.}
%     \label{fig:lowestvers}
% \end{figure}

%\begin{figure}
%\centering
 %   \includegraphics[width=9cm,height=9cm]{Figures/barchart-genre.png}
 %   \caption{The top represented genres in the data set, 86\% of the apps fall into these genres.}
 %   \label{fig:bargenre}
%\end{figure}

\subsection{Data preparation}
\setlength{\unitlength}{1in}%
\setlength{\fboxsep}{0pt}
\setlength{\fboxrule}{0pt}
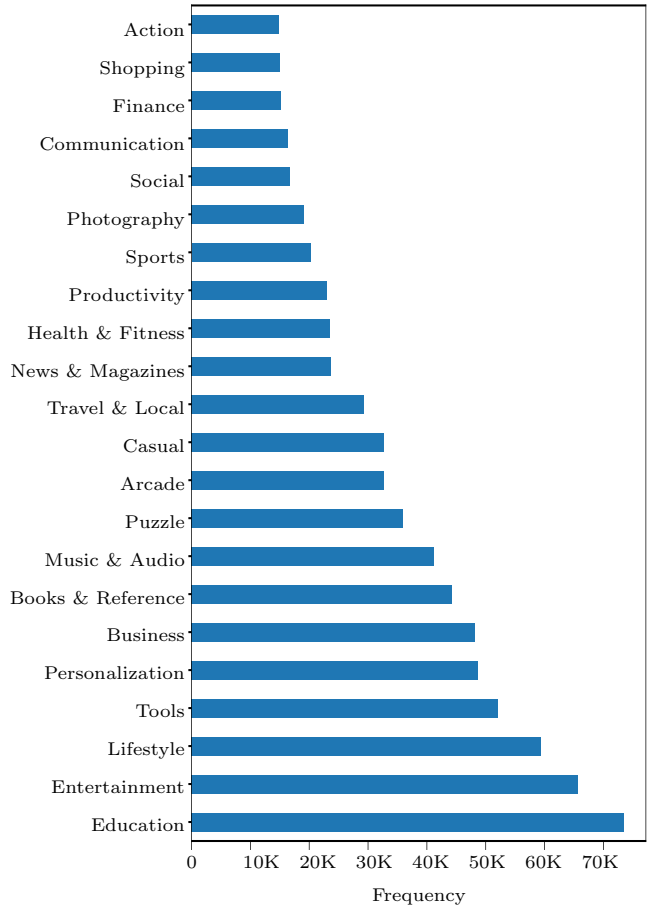
\begin{figure}[t]
    \scriptsize
    % \centering
    \vspace{-0.1in}
    \hspace{0.1in}
    \fbox{
        \begin{picture}(3,5)%
        \put(.15,0){\includegraphics[width=3in,height=5in]{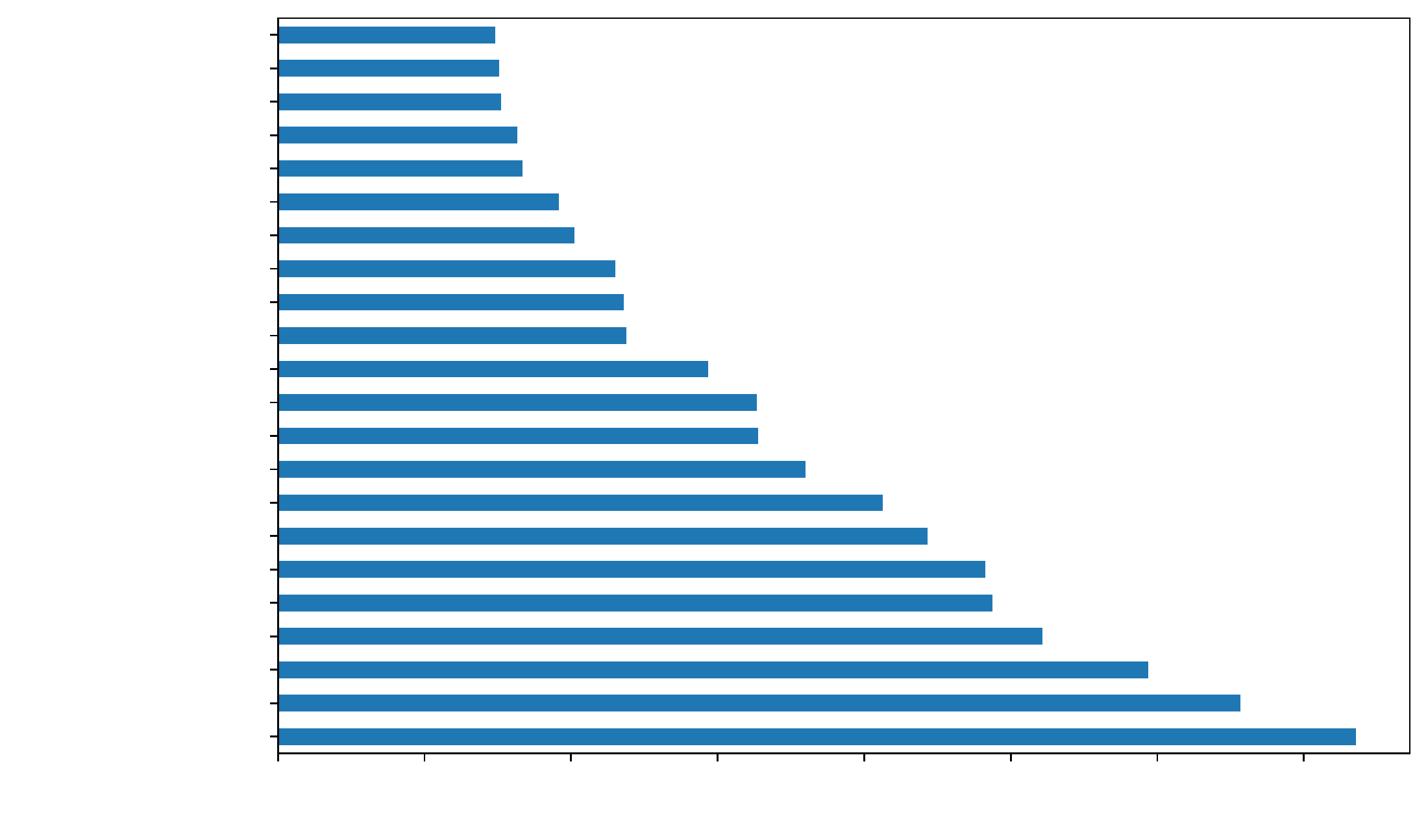}}%
        \put(.675,.2){
            \fbox{
                \begin{minipage}[t][0.1in]{2.35in}
                    \centering Frequency
                \end{minipage}
                }
            }
        \put(.67,.3){
            \rotatebox{90}{
                \fbox{
                    \begin{minipage}[t][1.45in]{0.1cm}
                        \setstretch{2.67}
                        \rotatebox{270}{0} \\
                        \rotatebox{270}{10K} \\
                        \rotatebox{270}{20K} \\
                        \rotatebox{270}{30K} \\
                        \rotatebox{270}{40K} \\
                        \rotatebox{270}{50K} \\
                        \rotatebox{270}{60K} \\
                        \rotatebox{270}{70K} \\
                    \end{minipage}
                }
            }
        }
        
        \put(-.375,5.08){
            % \rotatebox{90}{
            \fbox{
                \begin{minipage}[t][5in]{1in}\
                    \setstretch{1.79}
                    \vspace{0.04in}
                    \begin{flushright}
                        Action \\ 
                        Shopping \\ 
                        Finance \\
                        Communication \\
                        Social \\
                        Photography \\
                        Sports \\
                        Productivity \\
                        Health \& Fitness \\ 
                        News \& Magazines\\ 
                        Travel \& Local \\
                        Casual \\
                        Arcade\\
                        Puzzle\\
                        Music \& Audio \\
                        Books \& Reference \\
                        Business \\
                        Personalization \\
                        Tools \\
                        Lifestyle \\
                        Entertainment\\
                        Education
                    \end{flushright}
                \end{minipage}
            }
        }    
        \end{picture}%
    }
    \vspace{-0.2in}
    \caption{The top represented genres in the data set, 86\% of the apps fall into these genres.\vspace{-0.165in}}
   \label{fig:bargenre}
\end{figure}

\setlength{\unitlength}{1in}%
\setlength{\fboxsep}{0pt}
\begin{figure}[t]
    \scriptsize
    \centering
    % \fbox{
    \begin{picture}(3.25,2.4)
    \put(0.5,0.6){\includegraphics[width=2.75in]{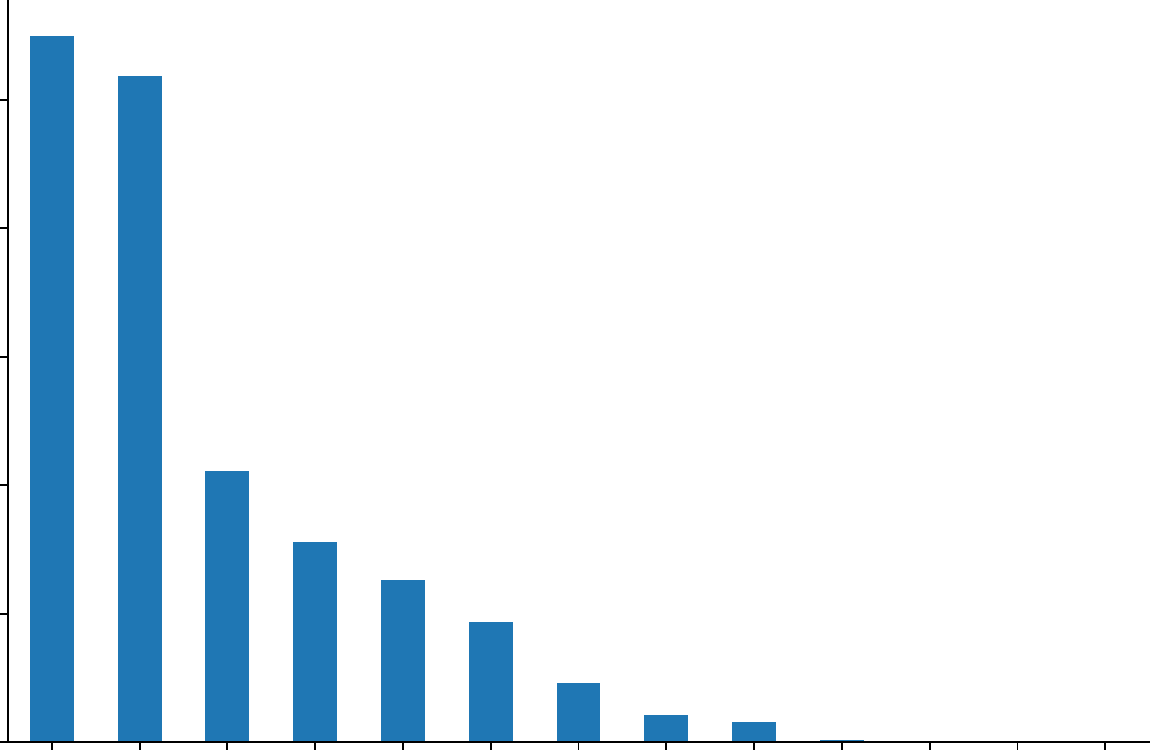}}
    \put(0.55,-0.46){
        \rotatebox{90}{
            \begin{minipage}[t][2.7in]{1in}
                \setstretch{1.9}
                \begin{flushright}
                Gingerbread \\
                Ice Cream Sandwich \\
                Froyo \\
                Jelly Bean \\
                Honeycomb \\
                Eclair \\
                Donut \\
                Cupcake \\
                Varies with device \\
                1.0\\
                1.1\\
                KitKat\\
                Lollipop\\
                \end{flushright}
            \end{minipage}
        }
    }
    \put(-0.05,2.13){
        \begin{minipage}[t][1.6in]{0.5in}
            \setstretch{2.775}
            \begin{flushright}
            250000\\
            200000\\
            150000\\
            100000\\
            50000\\
            0\\
            \end{flushright}                
        \end{minipage}
    }    
    \put(-0.025,0.6){
        \rotatebox{90}{
            \begin{minipage}[t][0.1in]{1.65in}
                \begin{center}
                    Frequency
                \end{center}
            \end{minipage}
        }
    }
    \put(0.5,-0.4){
        \begin{minipage}[t][0.1in]{2.65in}
            \begin{center}
                Lowest Android Version
            \end{center}
        \end{minipage}
    }
    \end{picture}
    \vspace{0.4in}
    \caption{A histogram that shows the distribution of the lowest Android version that was specified in all apps.}
    \label{fig:lowestvers}
\end{figure}

Since the data set was collected in different phases, the first step was to merge them together, hence the \emph{Merge} step in Fig.~\ref{fig:dscollection}. The merge is meant to keep all the apps with complete profiles, which entails the information from the Play store pages, APK files, and the three different status checks. This resulted in a total of 999, 530 applications as per Table~\ref{table:status}. We then excluded the applications that either had missing values due to crawling errors, or their manifest files could not be retrieved from their APK files. This step resulted in reducing the number of applications to 870,515, which we call the \emph{merged} data set. %Next, we excluded the applications that we could not retrieve their Manifest files from their APK files.
%. Finally, we removed any redundancy that might occur during any of the explained phases. 
%The result was a data set of 870,515 applications, which we call the \emph{merged} data set.

In Fig.~\ref{fig:bargenre}, we show the top 22 represented genres in the \textit{merged} data set (there are 48 distinct genres), nearly 86\% of the entire data set fall into these genres. The mean and standard deviation of the reviews for all apps are 3.4 and 1.7, respectively.
%The five-number summary of the reviews averages for all apps in the data set is as follows: min:0.00, 25\%:3.10, 50\%, 4.00, 75\%: 4.60, and max: 5.00. 
This suggests that the majority of the collected applications have in fact high review averages including the ones that were removed from the market. In fact, the mean for the reviews of the removed applications is 3.3 with a standard deviation of 1.77, in comparison to 3.6 and 1.77 for the applications that were kept in the market.  
%This includes the removed applications with a reviews average of 3.3 in comparison to 3.6 for the none removed. 
%In Figure~\ref{fig:reviewsavg}, we show the distribution of the applications' reviews average. 
Finally, in Fig.~\ref{fig:lowestvers}, we show the distribution of the lowest Android version of all applications in the data set. Nearly 60\% of the apps have \emph{Gingerbread} and \emph{Ice Cream Sandwich} as their lower Android version. In Section~\ref{sec:experiments} we describe the contribution of this information, lower Android version, in predicting the faith of an app.
%\subsection{Limitations and Challenges}
%\label{sec:limitations}
%First, our hypothesis is that the vast majority of the removed apps were bad apps, do not follow Google's policies and guidelines \footnote{https://play.google.com/about/developer-content-policy/}, while Google Play could
%remove apps with a number of other reasons \cite{RmvDroid}.

\subsection{Feature Engineering}
In this phase, we determined the most suitable features, also known as independent variables, and whether they require any further preprocessing.
% instance, the \emph{Description} feature is a free long text whose content requires sophisticated natural language processing to be analysed. In order to keep it simple, we decided to only use its length as a feature. We also excluded applications that are not in English.
We applied various techniques for the purpose of normalizing and standardizing all features. We call this collection the \emph{final data set}.

%\subsection{Features}
%\label{sect:features}
%three types of features: stemmed from the meta data of the apps, stemmed from the apk files (configuration), and computed.
%The cleaning, merging, and validating of the seed, apks, and status data sets resulted in a final data set of 870515 applications with 48 features.
In Table~\ref{table:allfeatures}, we list all features in the final data set, their sources, types, and the pre-processing operations that we applied to them. Twenty seven features came directly or indirectly from the \emph{seed} data set, and twenty features came from the \emph{APKs} data set. Seven of these features are of categorical type and the remaining are numerical. We applied the one hot encoding to all categorical features. 

\subsubsection{Input features of the Seed data set}
\label{sec:seedfeatures}
Table~\ref{table:contextfeatures} shows the list of 24 features of the \emph{Seed data set}, which we were able to crawl from the Google Play store pages for each application. The table contains real values for the features of the \emph{WhatsApp} Android application as an example. In Table~\ref{table:allfeatures} we show the 27 features (the rows where the Source column has Seed for its value) that were generated from these 24 variables. 

\renewcommand{\arraystretch}{0.65}
\begin{table}[H]
\footnotesize
\caption{The list of the 48 features that are used to build the predictive models. The abbreviation ``Cat." in the Type column stands for categorical. The ``Source" indicates the source data set as per Section~\ref{sec:dataset}. The ``Operation" column states the pre-processing that we have applied to each feature. The ``Transformed" string implies some kind of processing to the respective feature, for example, ``LenTitle" is obtained by measuring the length of the app's title. Finally, the highlighted rows indicate the features that were excluded when building the developer-centered model.\vspace{-0.2in}}
\label{table:allfeatures}
\center
\begin{tabular}{lllll}
\toprule
%\hline
\textbf{} & \textbf{Feature}          & \textbf{Type} & \textbf{Source} & \textbf{Operation} \\ 
\midrule 
%\hline
1               & Status                    & Target  & Status          & Aggregated          \\ %\midrule
\rowcolor{lightgray}
2               & OneStarRatings            & Int           & Seed            & Normalized               \\ %%\midrule
\rowcolor{lightgray}
3               & TwoStarRatings            & Int           & Seed            & Normalized               \\ %\midrule
\rowcolor{lightgray}
4               & ThreeStarRatings          & Int           & Seed            & Normalized               \\ %\midrule
\rowcolor{lightgray}
5               & FourStarRatings           & Int           & Seed            & Normalized               \\ %\midrule
\rowcolor{lightgray}
6               & FiveStarRatings           & Int           & Seed            & Normalized               \\ %\midrule
\rowcolor{lightgray}
7              & ReviewsAverage            & Float           & Seed            & None               \\ %\midrule
8               & LenTitle                  & Int           & Seed            & Transformed    \\  
%\midrule
9              & LenDescription            & Int           & Seed            & Transformed                   \\ %\midrule
\rowcolor{lightgray}
10              & LenWhatsNew               & Int           & Seed            & Transformed                   \\ %\midrule
11              & DeveloperWebsite        & Int           & Seed            & Transformed                   \\ %\midrule
12              & DeveloperEmail          & Int           & Seed            & Transformed          \\ %\midrule
13              & DeveloperAddress        & Int           & Seed            & Transformed          \\ %\midrule
14              & PrivacyPolicyLink     & Int           & Seed            & Transformed          \\ %\midrule

15             & Paid                      & Int           & Seed            & None               \\ %\midrule
\rowcolor{lightgray}
16              & MaxDownloadsLog       & Int           & Seed            & Logarithmic               \\ %\midrule

17              & LowestAndroidVersion  & Cat.   & Seed            & Derived            \\ %\midrule
18              & HighestAndroidVersion & Cat.   & Seed            & Derived            \\ %\midrule
19              & AndroidVersion            & Cat.   & Seed            & Encoded            \\ %\midrule
20              & DevRegisteredDomain       & Int           & Seed            & Transformed          \\ %\midrule 
\rowcolor{lightgray}
21            & DaysSinceLastUpdate & Int           & Seed          & Derived   \\ %\midrule
\rowcolor{lightgray}
22              & LastUpdated               & Int           & Seed            & Transformed          \\ %\midrule
23              & FileSize                & Int           & Seed            & Encoded               \\ %\midrule
24             & CurrentVersion            & Cat.   & Seed            & Transformed            \\ %\midrule
25              & Genre                     & Cat.   & Seed            & Encoded            \\ %\midrule
26              & ContentRating             & Cat.   & Seed            & Encoded            \\ %\midrule
27              & DeveloperCategory         & Cat.   & Seed            & Generated \& Encoded            \\ %\midrule
28              & IsSpamming                & Int           & Seed            & Generated          \\ %\midrule
29               & Storage                   & Int           & APKs            & Transformed          \\ %\midrule
30               & Calendar                  & Int           & APKs            & Transformed          \\ %\midrule
31              & Camera                    & Int           & APKs            & Transformed          \\ %\midrule
32              & Contacts                  & Int           & APKs            & Transformed          \\ %\midrule
33              & Location                  & Int           & APKs            & Transformed          \\ %\midrule
34              & Microphone                & Int           & APKs            & Transformed          \\ %\midrule
35              & Phone                     & Int           & APKs            & Transformed          \\ %\midrule
36              & Sensors                   & Int           & APKs            & Transformed          \\ %\midrule
37              & SMS                       & Int           & APKs            & Transformed          \\ %\midrule
38              & Net                       & Int           & APKs            & Transformed          \\ %\midrule
39              & Intent                    & Int           & APks            & Transformed          \\ %\midrule
40              & Bluetooth                 & Int           & APKs            & Transformed          \\ %\midrule
41              & App                       & Int           & APKs            & Transformed          \\ %\midrule
42              & Provider                  & Int           & APKs            & Transformed          \\ %\midrule
43              & Speech                    & Int           & APKs            & Transformed          \\ %\midrule
44              & NFC                       & Int           & APKs            & Transformed          \\ %\midrule
45              & Media                     & Int           & APKs            & Transformed          \\ %\midrule
46              & Hardware                  & Int           & APKs            & Transformed          \\ %\midrule
47             & Google                    & Int           & APKs            & Transformed          \\
% \midrule
48              & OS                        & Int           & APKs            & Transformed          \\ \hline
\end{tabular}
\end{table}

\begin{table}[h]
\center
\scriptsize
\caption{The features obtained from the Google Play store page for the \emph{WhatsApp} application as at February 2021.}
%\resizebox{\columnwidth}{!}{
\begin{tabular}{lllp{6.3cm}}
\toprule
% \hline
\textbf{} &\textbf{Feature} & \textbf{Type} & \textbf{Sample values} \\
\midrule 
\textbf{1}                   & \textbf{Description}                   & Text 
& WhatsApp Messenger is a FREE messaging app available for Android and other smartphones.... \\
%& \begin{tabular}[c]{@{}l@{}}WhatsApp Messenger is a FREE messaging app available \\ for Android and other smartphones....\end{tabular}                                   
\midrule
\textbf{2}                   &\textbf{Title}                         & Text                               & WhatsApp Messenger                                                                                                                                                      \\ \midrule
\textbf{3}                   &\textbf{Last Updated}                  & Date                               & May 13, 2020                                                                                                                                                            \\ \midrule
\textbf{4}                   &\textbf{Whats New}                     & Text                               & Group video and voice calls now support up to 8 participants. All participants need to be on the latest version of WhatsApp. \\ \midrule
\textbf{5}                   &\textbf{Reviews Average}               & Number                             & 4.3                                                                                                                                                                     \\ \midrule
\textbf{6}                   &\textbf{Price}                         & Number                             & 0.0                                                                                                                                                                     \\ \midrule
\textbf{7}                   &\textbf{Ratings}                       & Number                             & 114,391,572                                                                                                                                                             \\ \midrule
\textbf{8}                   &\textbf{One Star Ratings}              & Number                             & 4,000,000                                                                                                                                                               \\ \midrule
\textbf{9}                   &\textbf{Two Star Ratings}              & Number                             & 2,000,000                                                                                                                                                               \\ \midrule
\textbf{10}                   &\textbf{Three Star Ratings}            & Number                             & 2,391,572                                                                                                                                                               \\ \midrule
\textbf{11}                   &\textbf{Four Star Ratings}             & Number                             & 6,000,000                                                                                                                                                               \\ \midrule
\textbf{12}                   &\textbf{Five Star Ratings}             & Number                             & 100,000,000                                                                                                                                                             \\ \midrule
\textbf{13}                   &\textbf{Privacy Policy Link}           & Text                               & http://www.whatsapp.com/legal/\#Privacy                                                                                                                                 \\ \midrule
\textbf{14}                   &\textbf{Genre}                         & Text                               & Communication                                                                                                                                                           \\ \midrule
\textbf{15}                   &\textbf{Url}                           & Text                               & \url{https://play.google.com/store/apps/details?id=com.whatsapp}                                                                               \\ \midrule
\textbf{16}                   &\textbf{Content Rating}                & Text                               & PEGI 3                                                                                                                                                                  \\ \midrule
\textbf{17}                   &\textbf{Current Version}               & Text                               & 2.20.157                                                                                                                                                                \\ \midrule
\textbf{18}                   &\textbf{Android Version}               & Text                               & 4.0.3 and up                                                                                                                                                            \\ \midrule
\textbf{19}                   &\textbf{Developer Email}               & Text                               & android@support.whatsapp.com                                                                                                                                            \\ \midrule
\textbf{20}                   &\textbf{Developer Website}             & Text                               & https://www.whatsapp.com/                                                                                                                                               \\ \midrule
\textbf{21}                   &\textbf{Developer Name}                & Text                               & WhatsApp Inc.                                                                                                                                                           \\ \midrule
\textbf{22}                   &\textbf{Developer Address}             & Text                               & 1601 Willow Road Menlo Park, CA 94025                                                                                                                                   \\ \midrule
\textbf{23}                   &\textbf{File Size}                     & Number                             & 28M                                                                                                                                                                     \\ \midrule
\textbf{24}                   &\textbf{Downloads}                     & Number                             & 5,000,000,000+                       \\                                                 \bottomrule
\end{tabular}
%}
\label{table:contextfeatures}
\end{table}

For instance, in Table~\ref{table:contextfeatures}, feature number 7, \emph{Ratings} illustrates the number of users who rated the app by giving it a score from 1 to 5. The feature was used to normalize features 8-12, in order to produce features 2-6 in Table~\ref{table:allfeatures}. The \emph{Ratings} was thus removed from the list.  %We normalized the values of the star ratings by dividing them by the value of \emph{Ratings}. 
%The \emph{Ratings} feature illustrates the number of users who rated the app by giving it a score from 1 to 5. 
%All of these features are are entirely concerned with the user ratings of the app on Google Play store.  
%The \emph{Reviews Average} feature is kept as it is, a float value with the range $[0,5]$.

%features  \emph{Ratings} featur Table~\ref{table:allfeatures}
%the features  \emph{Ratings, One Star Ratings, Two Star Ratings, Three Star Ratings, Four Star Ratings, Five Star Ratings, and Reviews Average}, are entirely concerned with the user ratings of the app on Google Play store. 

%The \emph{Ratings} feature shows the number of users who have rated the app by giving it a score from 1 to 5. The \emph{One Star Ratings} and \emph{Five Star Ratings} features show the number of users who have given the application a score of 1 and 5, respectively. As such, we normalized the values of the star ratings by dividing them by the \emph{Ratings}, and then removed the \textit{Ratings} feature. The \emph{Reviews Average} feature is kept as it is, a float value with the range $[0,5]$.
%, leaving the features at 23.
As to the \emph{Title, Description,} and \emph{Whats New} features, we only considered their overall length in characters. The variables \emph{Developer Website, Developer Email, Developer Address}, and \emph{Privacy Policy Link}, were treated as binary; either present (1) or absent (0). The \emph{Paid} binary feature is based on the \emph{Price} feature, where 0 means it is a free app, and 1 means otherwise. The \emph{Downloads} are originally given in ranges $[x,y]$, and decided to take the logarithm of the maximum; i.e. $\log(y)$. For instance, if the number of Downloads for an app is given as [5000,10000], then we take $\log(10000)$. 

%We display the set of pre-processing steps that we applied on all features in Table~\ref{table:transfencode}.

%The developer email, the \emph{developer\_email} feature, was transferred into a binary value with 1 meaning it is present and 0 otherwise. 
Some features were derived from existing ones, such as the \emph{HighestAndroidVersion} and \emph{LowestAndroidVersion}, both of them were derived from the \emph{AndroidVersion} feature. The \emph{DevRegisteredDomain} feature was derived from the \emph{DeveloperWebsite} feature. It states whether the developer of the app has her own domain name or not. Additionally, the \emph{DaysSinceLastUpdate} feature was derived from the \emph{LastUpdated} feature. It is a continuous integer that represents the number of days since each app was last updated in comparison to other apps in the data set and we computed it as follows. First, we get the most recent date from the data set, which would be the maximum date. Second, we calculate the \emph{DaysSinceLastUpdate} for each app as the number of days between its \emph{LastUpdated} and the maximum date. As a consequence,
at least one app will get a zero value; the app which was updated most recently. We kept the \emph{LastUpdated} feature as well, however, we only considered its year value in four digits. 
The size of an app is not explicitly mentioned in the Play store. There are normally two possible values for it, either varies with the device or unspecified. Notably, the majority of the apps have an unspecified size and only a tiny proportion has a variable size. Thus, we encoded these two values to create the \emph{FileSize} feature, where 0 means the size is unspecified, and 1 means it varies.
For the \emph{CurrentVersion}, we only considered the major version number. We also encoded the \emph{Genre} and the \emph{ContentRating} features using the one hot encoding. 

%has been transformed in the following way: normally, the version is in the format with several different numbers separated by dots (e.g. 123.456.789 ). After transformation, only the first (main) number from CurrentVersion string is kept, so in case of the example, this will be 123.

Lastly, we calculated two additional features, namely \emph{IsSpamming} and \emph{DeveloperCategory}, based on the previous work in \cite{10.1145/3038912.3052712}.
The former is a binary value calculated based on the number of apps a developer has and their download count, where 1 means that the developer is a spammer and 0 otherwise. The %\emph{DeveloperCategory} feature
latter relies on the number of apps a developer has in the store. Each developer would be assigned any of the following categories; Aggressive (more than
50 apps released), Active (10 to 50 apps released), Moderate (2 to 10 apps released), and Conservative (released only 1 app). Spamming developers are aggressive developers that do not have any app with over 1M downloads and with an average install number below 10k.

%\emph{isSpamming}, which 

%For the title of the app, we looked into the language it was written in We turned the title of into the language in which it was written then these values mapped to numeric value in the range between 0-55. The following figure shows the distribution of application where x-axis represents the language that used to write the application title and y-axis represents the number of applications.

\subsubsection{APKs Input Features}
\label{sec:seedfeatures}
The features that we obtained from the \emph{APKs} data set came from extracting and then parsing the applications' manifest files\footnote{Every app must include an AndroidManifest.xml file that contains essential information about the app.}. We mainly focused on two components; the permissions and the system actions. The number of unique permissions and system actions slightly vary from one Android distribution to another. In this work, however, we considered 176 unique permissions and 134 unique system actions. The Android system classifies these permissions into dangerous and normal types. Dangerous permissions allow mobile apps to access users' sensitive data such as contacts, SMS, and pictures. Therefore, Android mandates applications to get users' consent and approval to be able to use them. On the other hand, normal permissions are presumably less risky, thus apps can obtain them without involving the users. The Android system further puts dangerous permissions into 9 groups; Storage, Calendar, Camera, Contacts, Location, Microphone, Phone, Sensors, and SMS. Each of these groups contains one or more permissions. We created a feature per each permission group with a value of 1 or 0, in which 1 means that the app has requested at least one permission of that group and 0 otherwise. 

Android allows third-party applications to register for listening to various system's events, such as when a new SMS arrives, a new call is made, and when the battery is low. The system sends out a broadcast whenever any of these events occurs. An application needs to be pre-configured in order to be able to listen to some of these events by including the corresponding system actions. As far as the actions are concerned, there is no preexisting classification to them. Instead, we relied on the top package name, which resulted in having 11 distinct groups/features; \emph{Net}, \emph{Intent}, \emph{Bluetooth}, \emph{App}, \emph{Provider}, \emph{Speech}, \emph{NFC}, \emph{Media}, \emph{Hardware}, \emph{Google}, and \emph{OS}. 

\subsubsection{Dependent variable: Status}
After completing the crawling phase, which included downloading the corresponding APKs, we determined the status of all applications, by checking on three different occasions whether they are still in the market or not. The first check was done on December 2018, the second on February 2019, and the last was completed during the months of May and June 2019. 
%In Figure~\ref{fig:statdist}, 
In Table~\ref{table:status}, 
we show the percentages of the applications that  were never found in the market upon the three checks, stayed in the market the whole time, and the ones that changed their status throughout this period. We call the first group \emph{removed}, the second \emph{stable}, and the latter \emph{mix}.
%In Table~\ref{table:status} we show the number of apps in each group. 
The apps that fall in the \emph{removed} and \emph{stable} groups represent 91.6\% of the entire data set. As such, we decided to focus only on these two groups and ignore the other ones because the applications in those groups do not have sufficient samples in comparison to the first two. 

%\setlength{\unitlength}{1cm}
%\begin{figure}
%    \footnotesize
%    \centering
%    \begin{picture}(10,6)
%    \put(0,0){\includegraphics[height=6cm]{Figures/appstatus.png}}
    %\put(2.65,5.65){\textbf{Removed}}    
 %   \end{picture}
%    \caption{The status distribution of all apps. The R and S stand for Removed and Stable, respectively. }
%    \label{fig:statdist}
% \end{figure}

\begin{table}[h]
\center
\footnotesize
\caption{Availability of apps in the market. A value of 1 means the app was not in the market on the indicated date, 0 means otherwise. }
\label{table:status}
\begin{tabular}{ccccc}
%\hline
\toprule
\multicolumn{1}{l}{\textbf{Dec 18}} & \multicolumn{1}{l}{\textbf{Feb 19}} & \multicolumn{1}{l}{\textbf{May-June 19}} & \multicolumn{1}{l}{\textbf{\#Apps}} & \textbf{Portion (\%)} \\
\midrule
%\hline
\rowcolor{lightgray}
1                                     & 1                                    & 1                                         & 553395                               & 50.7      \\ %\hline
\rowcolor{lightgray}
0                                     & 0                                    & 0                                         & 446135                               & 40.9      \\ %\hline
0                                     & 1                                    & 1                                         & 38848                                & 3.6       \\ %\hline
0                                     & 1                                    & 0                                         & 31804                                & 2.9       \\ %\hline
1                                     & 0                                    & 0                                         & 9715                                 & 0.9       \\ %\hline
1                                     & 0                                    & 1                                         & 6507                                 & 0.6       \\ %\hline
0                                     & 0                                    & 1                                         & 3005                                 & 0.3       \\ %\hline
1                                     & 1                                    & 0                                         & 1075                                 & 0.1       \\ \midrule
\multicolumn{3}{r}{\textbf{Total}}                                                                                       & \multicolumn{1}{l}{1090484}         & 100       \\ \bottomrule
\end{tabular}
\end{table}

\subsection{Prediction Model}
\label{sec:model}
The machine learning algorithm chosen for this research is the Extreme Gradient Boosting of Decision Trees or XGBoost for short. This decision is motivated by its outstanding performance on various Kaggle\footnote{\url{www.kaggle.com}} benchmark data sets among others, its efficiency in learning and applying a model together with the ability in determining the relevance of each independent variable, which facilitates the interpretation of the pipeline \cite{Alsahaf2019,FARRUGIA2020113318,Lovdal2020}.

XGBoost is a supervised learning algorithm that predicts the target class through aggregating the decisions of a number of regression trees. It uses the gradient descent algorithm during learning for the minimization of the loss function when configuring new trees. For further technical details we refer to \cite{chenatal16}. 

In order to counter for the imbalance in the distribution of the two classes (\textit{removed} and \textit{stable}) we embed the XGBoost in a bootstrap aggregating (bagging) approach. This ensures that an XGBoost model is trained with a balanced data set, something that is desirable in machine learning in order to avoid any bias. In practice, the bagging approach requires the bootstrapping with replacement of balanced subsets and use them to train XGBoost models. For a given app the prediction is then calculated using the majority voting rule of all the classifications achieved by the participating XGBoost models.

\section{Experiments and Results}
\label{sec:experiments}
\begin{figure}[t]
    \centering
    \includegraphics[scale=3, width=\textwidth]{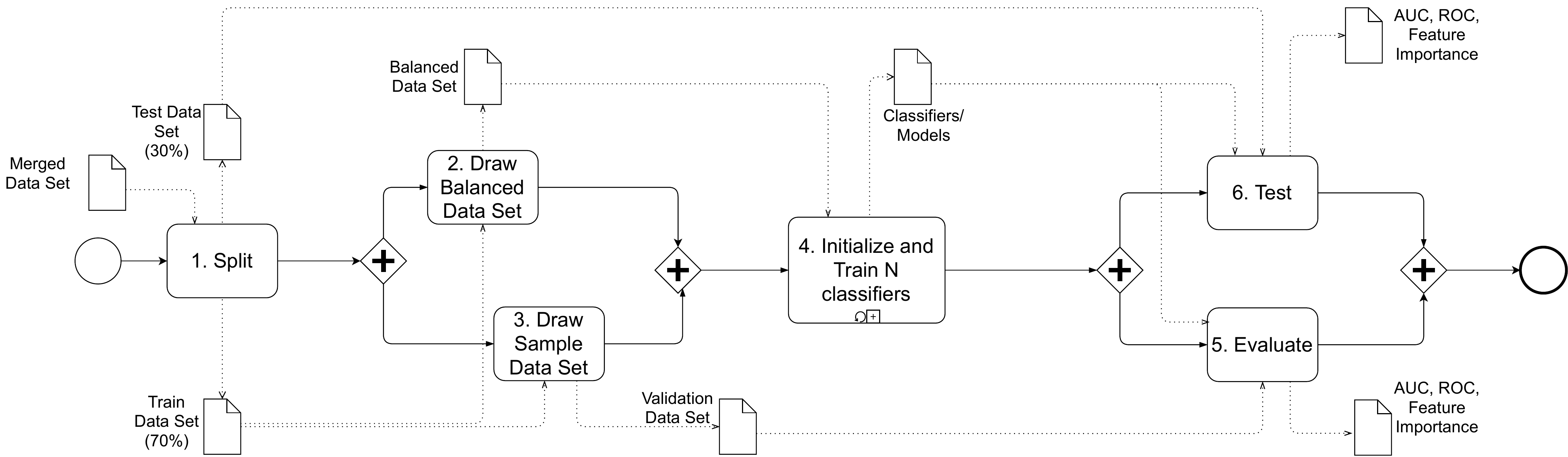}
    \caption{A high-level overview of the model building workflow in Business Process Modeling Notation (BPMN)}
    \label{fig:modeloverview}
\end{figure}

In Fig.~\ref{fig:modeloverview}, we show the steps that we have taken to build, train, and test the two models. In Step 1, the \emph{final data set} of $870, 515$ records is split (stratified random sampling) into training data (70\%: $609, 360$) and test data (30\%: $261, 155$). In Step 2, a balanced data set is drawn from the training data. We experimented with the following sizes of the balanced data set; $2K, 5K, 10K, 25K, 50K,$ and $100K$. Each balanced subset is determined by randomly selecting, with replacement, the same number of removed and not removed apps from all apps in the training data. In Step 3, a validation data set is drawn from the training data, which has the same class distribution and size as the test data. In step 4, a number of XGBoost classifiers are initialized and trained using a different balanced data set that is sampled as mentioned in Step 2. We experimented with the following number of classifiers: $3, 5, 7, 9, 11, 13$, and $15$. Note that we do not fine tune the involved models/classifiers. We configure them using default values for all parameters except for the number of trees $n\_trees$ and the maximum depth of each tree $max\_depth$. For these two parameters we pick a random value from the following predefined lists: $max\_depth \in \{2,3\}$ and $n\_trees \in \{256,512\}$. Such values ensure that the learned XGBoost models are not fine tuned on the training data as they consist of an ensemble of very shallow decision trees. This is important to avoid overfitting. A similar approach was adapted in \cite{Lovdal2020}. For a given app the prediction (removed or not removed) is then determined from the average score of all predictions by the participating XGBoost models. Steps 5 and 6 are used to estimate the performance of the validation and test sets in terms of ROC and AUC, as well as the feature importance scores. 

\subsection{User- and Developer-centered models}
\label{subsec:removedfeatures}
The above experiments were first conducted using the 47 features that are shown in Table~\ref{table:allfeatures}. The resulting predictive model that relies on the full set of 47 features is distinctively suitable to the end users, who may use it to choose applications that will most likely remain in the market. We refer to this model as \textit{user-centered}. The Google Play store may also use the resulting model to identify violating applications after they have been admitted into the store. 

However, in order to aid developers before submitting their applications to the market or to aid the Play store before admitting the applications, another model is needed. The new model, which we call \textit{developer-centered}, shall rely on less features, the ones that are only available before deployment. In Table~\ref{table:allfeatures} we indicate these features with the non-highlighted rows. Thus, the developer-centered model uses 37 features, as it excludes the following 10 features that are only available after deployment: \emph{OneStarRatings, TwoStarRatings, ThreeStarRatings, FourStarRatings, FiveStarRatings, ReviewsAverage, LenWhatsNew, MaxDownloadsLog, DaysSinceLastUpdate, LastUpdated}. 

% \subsection{Model 1: After Admission}
% \label{subsec:Model1}
In order to select the best user- and developer-centered models we generate the receiver operating characteristic (ROC) curves for different settings and choose the models that achieve the highest area under ROC curve (AUC). The ROC curve demonstrates how the true and false positive rates change as a function of the operating point. The operating point is the threshold (between 0 and 1) that determines the predicted label of the given sample. For instance, for an operating point of 0.5, if the average predicted score of all involved XGBoost classifiers is below 0.5 then the given sample is labelled as ``to be removed", otherwise ``to be kept". For each considered operating point we compute the confusion matrix that consists of the number of true positives (TP), false positives (FP), true negatives (TN) and false negatives (FN). The true positive rate (TPR) is then the ratio of TP and all positive predictions (TP + FP), while the false positive rate (FPR) is the ratio of FP and the total number of negatives (FP and TN). The AUC is the area under the ROC curve, which can have a value between 0 and 1. The higher (maximum 1) the AUC the better the performance of the respective model. An AUC of 0.5 for a two-class problem indicates that the model is not better than a random decision, and an AUC of 0 represents predictions that are completely opposite than desired.

\begin{figure*}[t]
    \footnotesize
    %\centering
    \begin{tabular}{@{}c@{}@{}c@{}}
        User-centered & Developer-centered \\
        \includegraphics[width=0.5\textwidth]{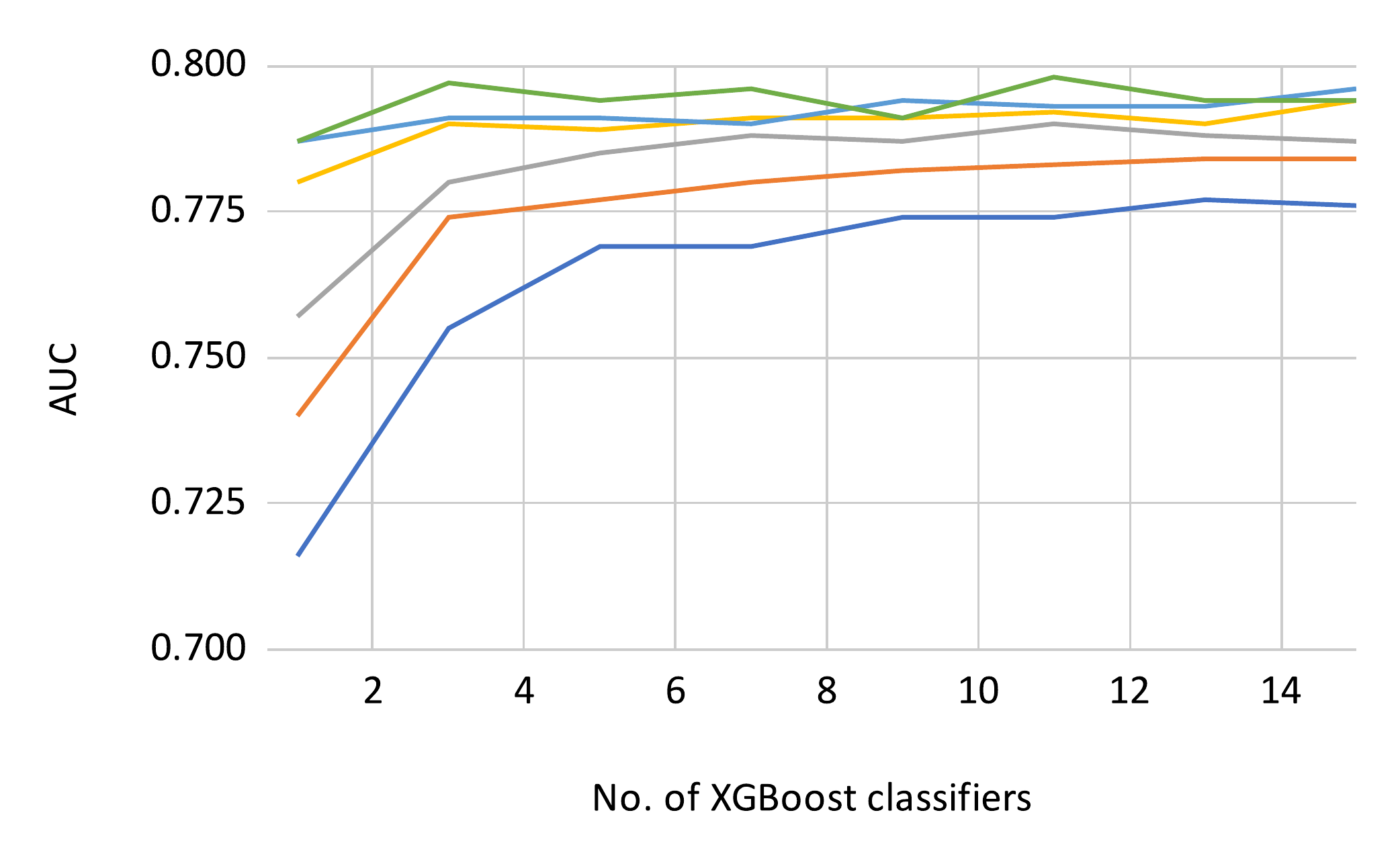} &
        \includegraphics[width=0.5\textwidth]{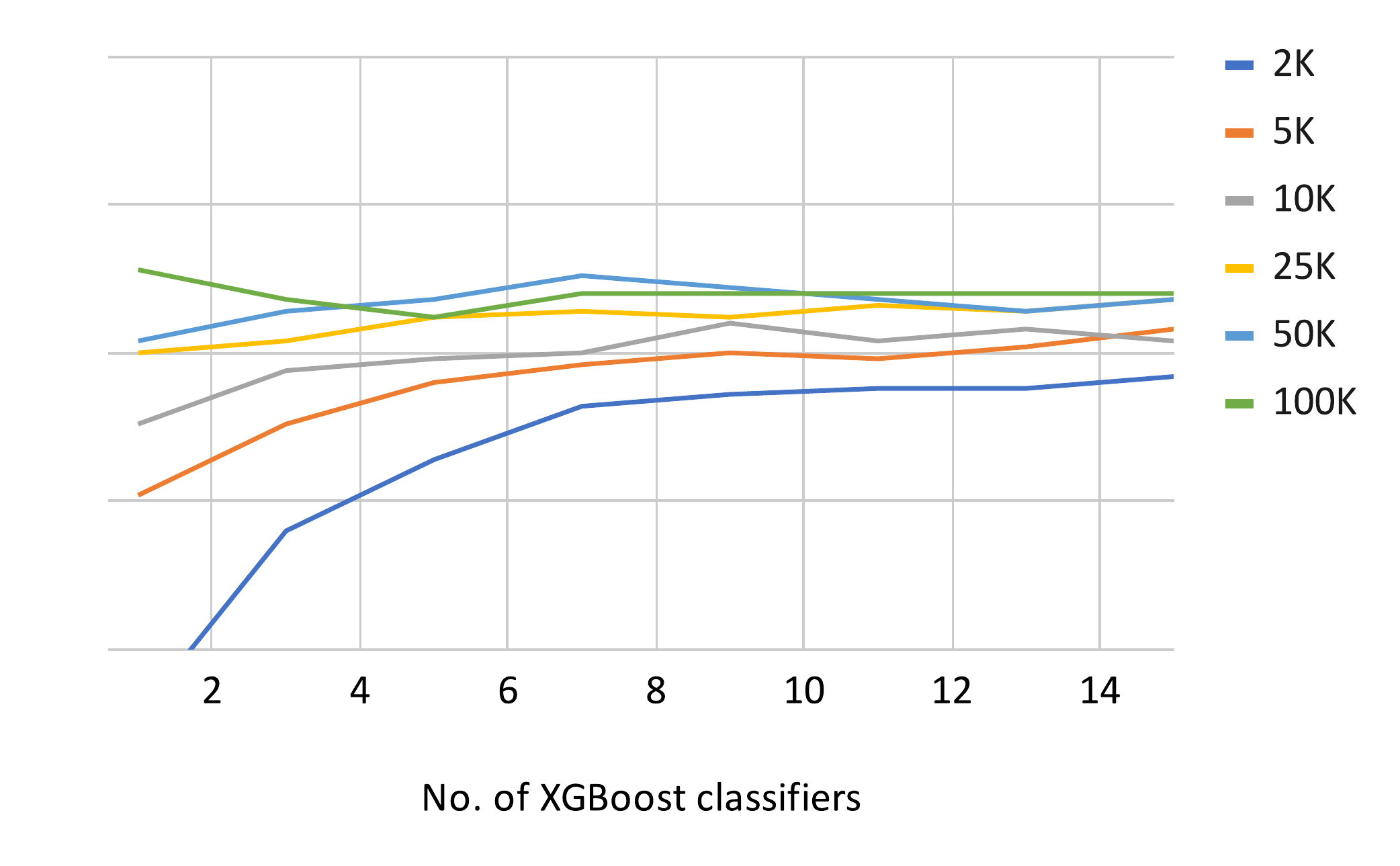} \vspace{-0.15in}
    \end{tabular}
    \caption{Evaluation in terms of Area under the ROC curves (AUC) of the (left) user-centered and the (right) developer-centered models on the validation data with respect to the number of XGBoost classifiers used and the sizes of the balanced data sets used to train such models. \vspace{-0.3in}}
    \label{fig:aucclssize}
\end{figure*}

In Fig.~\ref{fig:aucclssize} we show the AUC values for a number of experiments with varying number of XGBoost classifiers and sizes of the balanced training sets that are used to evaluate the user-centered and the developer-centered models on the validation set, respectively. The results show that the best performance of the user-centered model is achieved with 11 XGBoost classifiers, while for the developer-centered model the best performance is achieved with only 1 XGBoost classifier. For both models, the best performance is achieved with balanced training sets of $100K$ in size. 

In Fig.~\ref{fig:ROCcurves} we illustrate the ROC curves together with their AUC values of the selected models when applied to the test set. We also compute the AUC scores for the models when applied to the validation set. The high similarity between the AUCs of both models achieved for the validation (user-centered: 0.798, developer-centered: 0.764) and test sets (user-centered: 0.792, developer-centered: 0.762) demonstrate the generalisation ability of our approach.

%The average AUCs of	0.798 and 0.792 for the validation and test sets, respectively. 

%In Figure~\ref{fig:ROCcurves} we present the average ROC curves of our predictive model across the experiments that we obtained on the validation and the test sets.

%The average AUCs of 0.789 and 0.788 for the validation and test sets, respectively. The similar results demonstrate the generalisation ability of our approach.
 
% \begin{figure}[t]
%     \centering
%     \includegraphics[width=0.5\textwidth]{Figures/auc-validation-model2-1.pdf}
%     \caption{A line chart showing the AUC values for evaluating the \textit{Developer-centered Model} using different no. of XGBoost classifiers and different sizes of the balanced training sets.}
%     \label{fig:aucclssize-model2}
% \end{figure}

\setlength{\unitlength}{1in}%
\setlength{\fboxsep}{0pt}
\setlength{\fboxrule}{0pt}
\begin{figure}[t]
    \scriptsize
    \centering
    \fbox{
        \begin{minipage}[t][1.8in]{0.75\textwidth}
            \begin{center}
            \hspace{-0.2in}
            \fbox{
                \begin{picture}(3.05,1.8)
                \put(0.325,0.2)
    % {\includegraphics[width=2.75in]{Figures/roc-tests-2models.pdf}}
    {\includegraphics[width=2.75in]{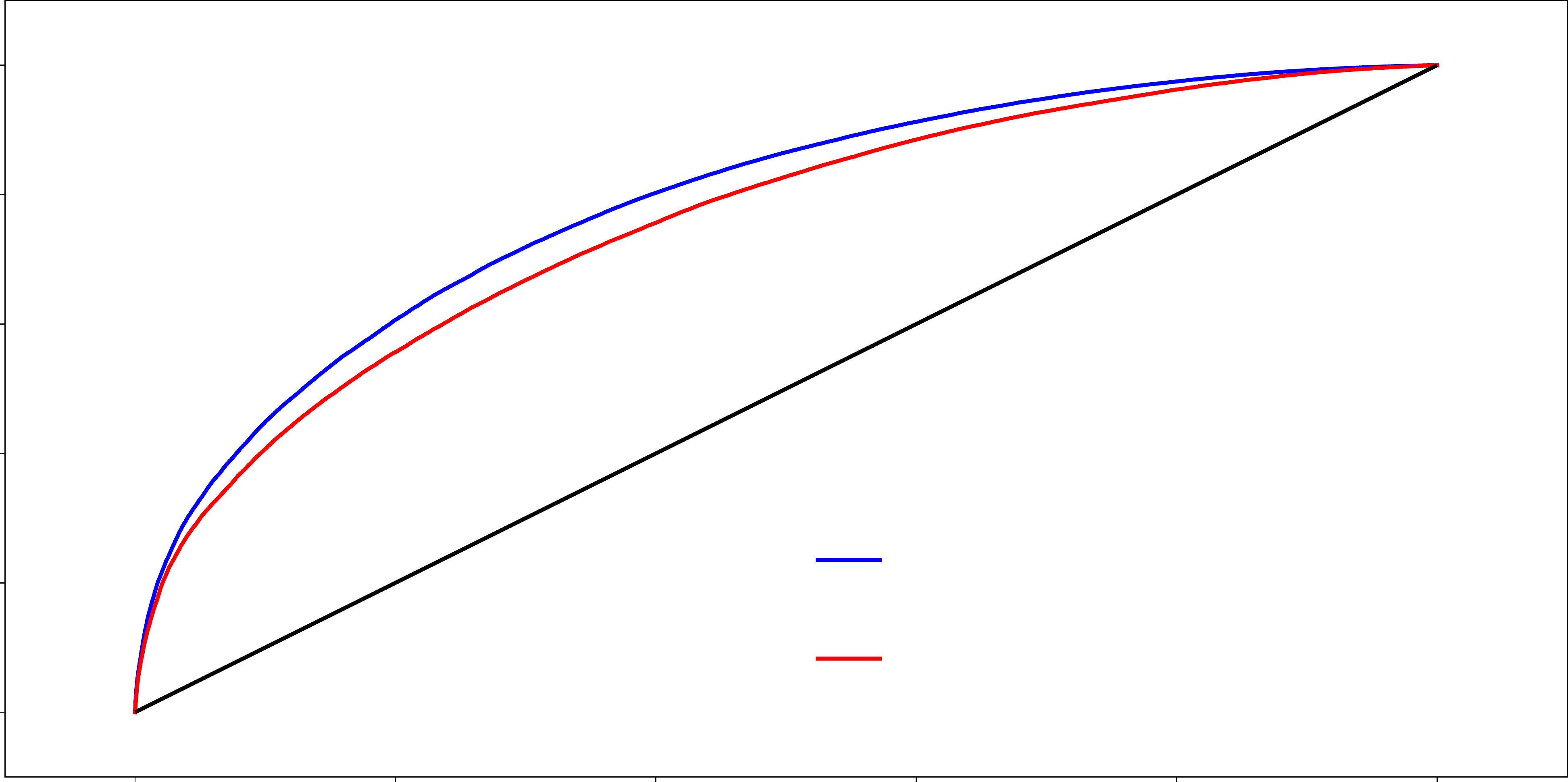}}
     %{\includegraphics[width=2.75in]{Figures/Fig9.pdf}}
                \put(0.47,0.01){
                    \rotatebox{90}{
                        \fbox{
                        \begin{minipage}[t][2.5in]{0.1in}
                            \setstretch{4.12}
                            \begin{flushright}
                                \rotatebox{270}{0.0}\\
                                \rotatebox{270}{0.2}\\
                                \rotatebox{270}{0.4}\\
                                \rotatebox{270}{0.6}\\
                                \rotatebox{270}{0.8}\\
                                \rotatebox{270}{1.0}\\ 
                            \end{flushright}
                        \end{minipage}
                        }
                    }
                }
                \put(0.09,1.65){
                    \fbox{
                    \begin{minipage}[t][1.21in]{0.13in}
                        \setstretch{2.05}
                        \begin{flushright}
                        1.0\\
                        0.8\\
                        0.6\\
                        0.4\\
                        0.2\\
                        0.0\\
                        \end{flushright}                
                    \end{minipage}
                    }
                }    
                \put(0,0.18){
                    \rotatebox{90}{
                        \fbox{
                        \begin{minipage}[t]{1.275in}
                            \begin{center}
                                True positive rate
                            \end{center}
                        \end{minipage}
                        }
                    }
                }
                \put(0.305,0){
                    \fbox{
                    \begin{minipage}[t]{2.68in}
                        \begin{center}
                            False positive rate
                        \end{center}
                    \end{minipage}
                    }
                }
                \put(1.88,0.56){AUC of Model 1~$=0.79$}
                \put(1.88,0.39){AUC of Model 2~$=0.76$}
                \end{picture}
            }
            \end{center}
        \end{minipage}
    }
    \caption{ROC curves obtained from the test data for Model 1 (user-centered) and Model 2 (developer-centered).}
    \label{fig:ROCcurves}
\end{figure}
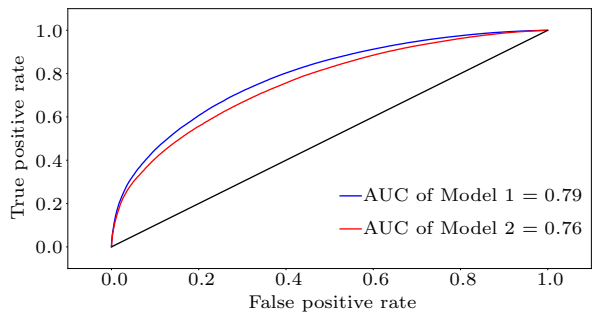

As a byproduct, the XGBoost models provide us with an importance score for each independent variable. Importance scores are calculated for all attributes in each decision tree. For each tree, an attribute that is used more often to make key decisions, is given a high importance score. Then, the final feature importance of a particular attribute is summed up and divided by the number of decision trees. 
%Figure~\ref{fig:importances} shows 
We determine the ranking of the features after averaging the importance scores across the participating XGBoost classification models. For the user-centered predictive model the \emph{ContentRating, PrivacyPolicyLink and DeveloperWebsite} are the top most important features in predicting the removability of an app. In Fig.~\ref{fig:featureimportance}, we show the averaged and normalized importance scores of the top 20 features across the 11 XGBoost classifiers. Similarly, in Fig.~\ref{fig:featureimportancem2} we show the normalized scores of the 20 most relevant features as achieved by the single XGBoost classifier that defines the developer-centered model. For this model the \emph{ContentRating\_Teen, PrivacyPolicyLink and IsSpamming} are the top most important features.
%in predicting the removability of an app. 
Both models share the top 5 most important features, which demonstrate their predictive power in both scenarios. Moreover, both models highlight the importance of including a \emph{PrivacyPolicyLink} upon submitting an app to the store. In addition, the difference in the number of features between the user- and developer-centered models (the user-centered model has 47 features, and the developer-centered model has 37 features) did not have a significant impact on the performance. 

%is crucial in keeping the 
% \subsection{Model 2: Before Admission}
% \label{subsec:Model1}
% In Figure~\ref{fig:aucclssize-model2}, we show the AUC values for a number of experiments with varying number of classifiers and sizes of the balanced data set that are used to evaluate the second model. 
% Our results showed that the optimal number of classifiers were 1, and the optimal size for the balanced data set was 100K. The average AUCs of	0.764 and 0.762 for the validation and test sets, respectively. 
% The similarity between the validation and testing results for both models demonstrate the generalisation ability of our approach. 

%It is worth mentioning here that out of those 10 features, 3 are seen in the top 20 list for the user-centered model (model 1), namely, \emph{MaxDownloadLog, DaysSinceLastUpdate, LenWhatsNew}. 

% In Figure~\ref{fig:tests} we present the average ROC curves for both predictive models across the experiments that we obtained on the test sets. 
% We further averaged and normalized the importance scores of the top 20 features for this model. Out of the 20 features, 11 were seen in the top 20 of model 1, Figure~\ref{fig:featureimportance}. 
% \begin{figure}
%     \centering
%     \includegraphics[width=0.5\textwidth]{Figures/15classifiers-AUC-val-tst.pdf}
%     \caption{ROC curves obtained from the validation and test data.}
%     \label{fig:auc}
% \end{figure}

\setlength{\unitlength}{1in}%wp
\setlength{\fboxsep}{0pt}
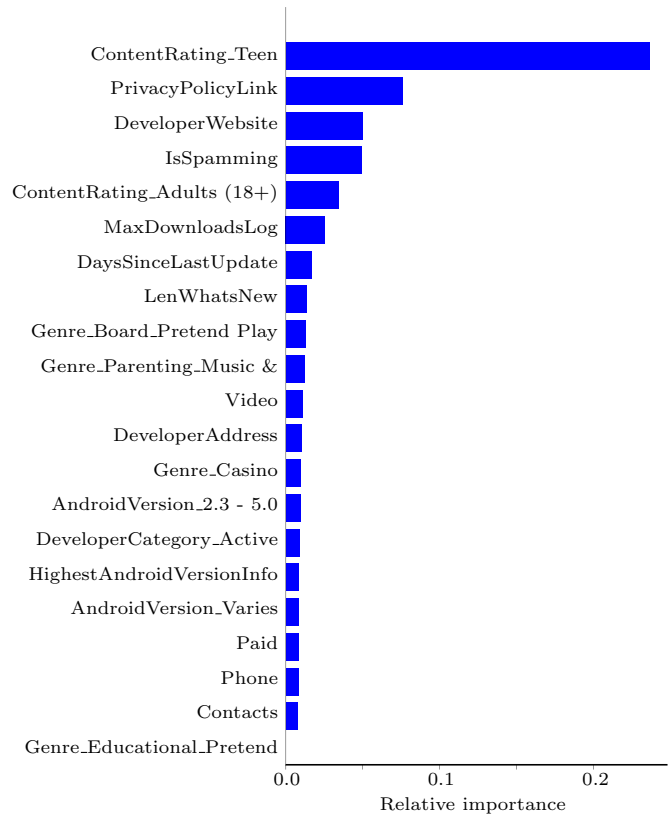
\begin{figure}[t]
    \scriptsize
    \centering
    % \fbox{
    \begin{picture}(3.25,4)%
    \put(1.25,0){\includegraphics[width=2in,height=4in]{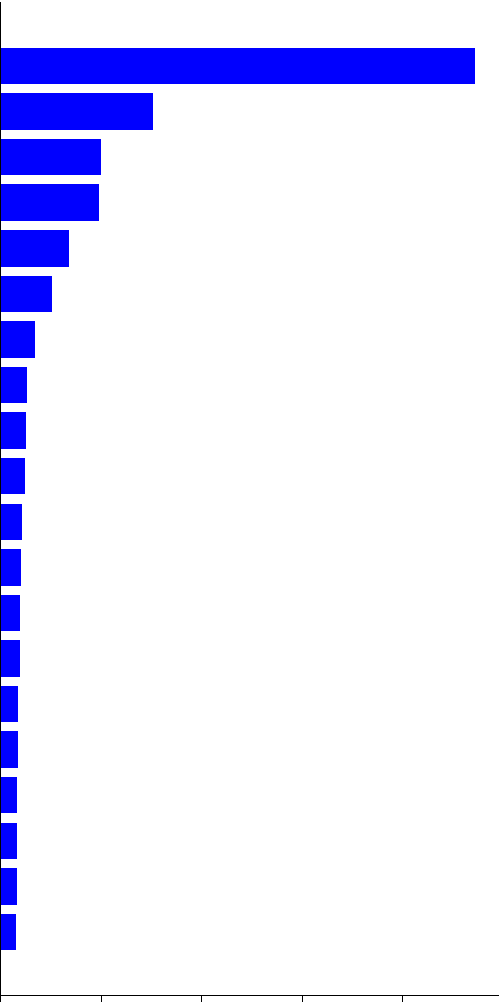}}%
    \put(1.22,-0.21){
        % \fbox{
        \begin{minipage}[t][0.1in]{1.95in}
            \begin{center}
                Relative importance
            \end{center}
        \end{minipage}
        }
        % }
    \put(1.15,-0.12){
        \rotatebox{90}{
            % \fbox{
                \begin{minipage}[t][1.5in]{0.1cm}
                    \setstretch{7.3}
                    \rotatebox{270}{0.0} \\
                    \rotatebox{270}{0.1} \\
                    \rotatebox{270}{0.2} \\
                \end{minipage}
            % }
        }
    }
    
    \put(-0.325,4.05){
        % \rotatebox{90}{
        % \fbox{
        \begin{minipage}[t][4in]{1.5in}\
            \setstretch{1.64}
            \vspace{0.04in}
            \begin{flushright}
                ContentRating\_Teen \\ 
                PrivacyPolicyLink \\ 
                DeveloperWebsite \\
                IsSpamming \\
                ContentRating\_Adults (18+) \\
                MaxDownloadsLog \\
                DaysSinceLastUpdate \\
                LenWhatsNew \\
                Genre\_Board\_Pretend Play \\
                Genre\_Parenting\_Music \& Video \\ 
                DeveloperAddress \\
               Genre\_Casino \\
               AndroidVersion\_2.3 - 5.0 \\
               DeveloperCategory\_Active \\
                HighestAndroidVersionInfo\\
                AndroidVersion\_Varies \\
                Paid \\
                Phone \\
                Contacts\\
                Genre\_Educational\_Pretend
            \end{flushright}
        \end{minipage}
        % }
    }    
    \end{picture}%
    % }
    \vspace{0.2in}
    \caption{Averaged and normalized importance scores of the top 20 features across the 11 participating XGBoost classifiers in the user-centered predictive model. The features names containing the character ``\_" are dummy variables created from the encoding of the categorical features listed in Table~\ref{table:allfeatures}.}
    \label{fig:featureimportance}
\end{figure}

\setlength{\unitlength}{1in}%
\setlength{\fboxsep}{0pt}
\begin{figure}[t]
    \scriptsize
    \centering
    % \fbox{
    \begin{picture}(3.25,4)%
    \put(1.25,0){\includegraphics[width=2in,height=4in]{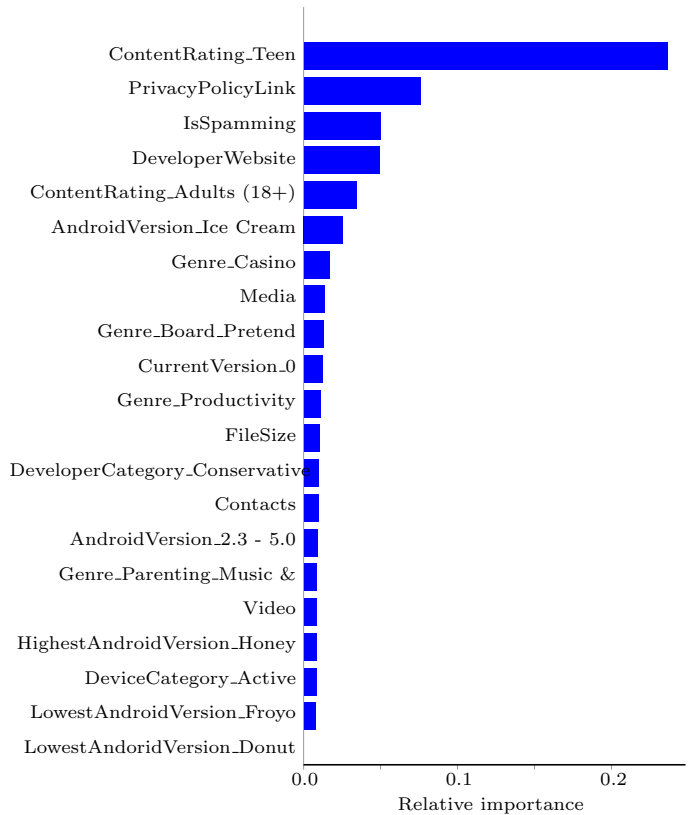}}%
    \put(1.22,-0.21){
        % \fbox{
        \begin{minipage}[t][0.1in]{1.95in}
            \begin{center}
                Relative importance
            \end{center}
        \end{minipage}
        }
        % }
    \put(1.15,-0.12){
        \rotatebox{90}{
            % \fbox{
                \begin{minipage}[t][1.5in]{0.1cm}
                    \setstretch{7.3}
                    \rotatebox{270}{0.0} \\
                    \rotatebox{270}{0.1} \\
                    \rotatebox{270}{0.2} \\
                \end{minipage}
            % }
        }
    }
    
    \put(-0.325,4.05){
        % \rotatebox{90}{
        % \fbox{
        \begin{minipage}[t][4in]{1.5in}\
            \setstretch{1.64}
            \vspace{0.04in}
            \begin{flushright}
                ContentRating\_Teen \\ 
                PrivacyPolicyLink \\ 
                IsSpamming \\
                DeveloperWebsite \\
                ContentRating\_Adults (18+) \\
                AndroidVersion\_Ice Cream \\
                Genre\_Casino \\
                Media \\
                Genre\_Board\_Pretend \\
                CurrentVersion\_0 \\
                Genre\_Productivity \\
                FileSize \\
                DeveloperCategory\_Conservative \\
                Contacts \\
                AndroidVersion\_2.3 - 5.0 \\
                 Genre\_Parenting\_Music \& Video \\ 
                  HighestAndroidVersion\_Honey \\
                  DeviceCategory\_Active \\
                  LowestAndroidVersion\_Froyo \\
                  LowestAndoridVersion\_Donut 
                
            \end{flushright}
        \end{minipage}
        % }
    }    
    \end{picture}%
    % }
    \vspace{0.2in}
    \caption{Averaged and normalized importance scores of the top 20 features of the single XGBoost classifier in the developer-centered predictive model. The features names containing the character ``\_" are dummy variables created from the encoding of the categorical features listed in Table~\ref{table:allfeatures}.}
    \label{fig:featureimportancem2}
\end{figure}

% \subsection{Time Difference}
% \label{subsec:time}
% We also looked at the time needed to generate the final predictive model with different number of features and sizes of the balanced data set. We measured the time for 2X6 = 12 different settings: number of features either 47 or 37, and the size of the balanced data set was 2K, 5K, 10K, 25K, 50K and 100K.

\section{Discussion}
\label{sec:discussion}
%In this section, we will first discuss the results of the first set of experiments that we conducted to create the first predictive model, we then move the discussion into the second model.

% \subsection{Time Analysis}
% \label{subsec:timeanalysis}
% In Figure~\ref{fig:timediff}, we show the comparison between the time taken to generate the two models using 6 different sizes for the balanced data set. As the figure shows, the difference is more significant with smaller balanced data set than larger data sets. 
% One explanation would be that the removed features were all integers, thus, no encoding was required. The total number of features after encoding for the full data set was 2401, in comparison to 2391 for the less.

% \begin{figure}
%     \centering
%     \includegraphics[width=0.5\textwidth]{Figures/timecomp1.pdf}
%     \caption{The difference in the time (minutes) needed to generate the two models.}
%     \label{fig:timediff}
% \end{figure}

%To the best of our knowledge we are the first to 
In this work we proposed two predictive models that can indicate whether an app will stay in the Play store or eventually get removed by Google.  Our results suggest that such a model can be built with a decent accuracy by largely relying on the app meta data that are publicly available and a number of elements that can be easily extracted from an app's manifest file. The AUCs of 0.792 and 0.762 for the user- and developer-centered models, respectively, represent the probabilities that they achieve higher scores for apps that will eventually be removed by the app store compared to those that will be retained. Therefore, the closer such an AUC (or probability) is to 1 the more accurate the model is. Typically, AUCs higher than 0.7 indicate strong effects between the independent and dependent variables \cite{Rice2005}. In practice, we would need to determine a threshold between 0 and 1 such that an app that results in a value above the threshold by our respective predictive model will be considered ``to be removed", otherwise ``to be kept". This threshold can be set to be the one that yields the maximum harmonic mean (or \textit{F}-score) of precision and recall on a validation set. Such a validation set can be drawn from the training set with the same size and class distribution as the test set (see Section~\ref{sec:experiments}).

We identified the need for both the user-centered and the developer-centered models, as 
%We believe that our model 
they would be practically useful for users, developers and app stores alike. The user-centered model will guide users into installing applications that are more likely to stay in the store. Developers can utilize the corresponding model to predict whether their applications are prone to be removed or not. Finally, online app stores will be able to use both models to filter out violating applications upon admission or thereafter.

As a result, we identify two deployment scenarios.
%\subsection{Deployment}
%\label{subsec:implementation}
%We envision different scenarios for using our predictive models in real life implementations. 
First, the user-centered model can be incorporated into a mobile app or a browser's extension/plugin. The former can work in two modes: before installing an application from the designated market or after. The user can use the latter on a desktop computer while browsing app stores to choose a new mobile app. Second, the developer-centered can be incorporated into a website that developers can use before submitting an application to the online store. 
%Unlike users, developers will be required to fill out their apps' information about their apps. 
%implementation mode
%user: mobile app or a browser plug in
%developer: a web page
%app store: back-end, thus, they can use all the features
% Almost all features can be obtained from the google play store with the exception of broadcast receivers
\subsection{Threats to Validity}
\label{subsec:validity}
In this work, the collection of the data set from the Google Play store is done under two main assumptions. The first assumption states that all applications in the store are benign \cite{PlayProtect2020}. The second one states that the only reason applications were  removed/disappeared from the store while crawling was due to their violations of the store's policy and recommendations \cite{8595206}. However, there could be other reasons, for example, removing applications by their owners temporally or permanently, which we do not take into account. While here we used the Google Play store as a case study, the proposed approach is also applicable to the Apple store and to other ones that share similar properties of mobile applications.
\section{Conclusion and future work}
\label{sec:conclusion}
The rapid increase of low quality and/or violating apps in the online stores has provoked stores' maintainers into employing strict measures. As a result, large number of apps are continuously removed from the stores. Removing mobile applications after they have been admitted into online stores negatively affect the experience of end users and the reputation of app developers. Thus, in this work, we propose two predictive models, which we call user- and developer-centered. The former aids mobile users and app stores to determine the future of the app \textit{after} deployment, while the latter supports developers and app stores \textit{before} deployment. Our models consider the meta data of an app that is publicly available on the play store. Additionally, they incorporate the permissions that the app requests in its manifest file and the system actions that it is registered to listen to. 
%To the best of our knowledge we are the first to propose such a data-driven approach.
%that can indicate whether an app will be accepted or not by the store. 

Both models were trained and validated using a very large data set of apps that we collected from the Google Play store. The data set is made publicly available
%\footnote{
 \cite{H0YJFT_2022}.
%is set here: https://cutt.ly/DQ0ib8W. If we have the honour to have this paper accepted for publication, we will move it to \url{https://dataverse.nl} so that it can be used for benchmark purposes in future investigations.} 
%We generated a very large data set for the evaluation of our approach, which we will make publicly available. We hope that our data set and encouraging results will serve as a basis for further investigations.

In future, we firstly plan to investigate more sophisticated methods to extract information from unstructured text-based variables such as the \emph{Description}, \emph{WhatsNew}, and \emph{PrivacyPolicyLink} features.
%. Example, instead of simply extracting the length of the \emph{Description}, Natural Language Processing techniques can be used to extract the most important keywords and then encode them in a vector space. This applies to other similar variables. 
%Another direction for future work would be to investigate. 
Secondly, we will look into adapting the proposed data-driven approach to work with other app stores, such as Apple and beyond \cite{10.1145/3278532.3278558}.

%\section*{References}

\bibliography{mybibfile}

\begin{thebibliography}{10}
\expandafter\ifx\csname url\endcsname\relax
  \def\url#1{\texttt{#1}}\fi
\expandafter\ifx\csname urlprefix\endcsname\relax\def\urlprefix{URL }\fi
\expandafter\ifx\csname href\endcsname\relax
  \def\href#1#2{#2} \def\path#1{#1}\fi

\bibitem{8587187}
K.~O. Elish, H.~Cai, D.~Barton, D.~Yao, B.~G. Ryder, Identifying mobile
  inter-app communication risks, IEEE Transactions on Mobile Computing 19~(1)
  (2020) 90--102.
\newblock \href {https://doi.org/10.1109/TMC.2018.2889495}
  {\path{doi:10.1109/TMC.2018.2889495}}.

\bibitem{8714014}
C.~Bi, G.~Xing, T.~Hao, J.~Huh-Yoo, W.~Peng, M.~Ma, X.~Chang, Familylog:
  Monitoring family mealtime activities by mobile devices, IEEE Transactions on
  Mobile Computing 19~(8) (2020) 1818--1830.
\newblock \href {https://doi.org/10.1109/TMC.2019.2916357}
  {\path{doi:10.1109/TMC.2019.2916357}}.

\bibitem{7173047}
W.~Gu, L.~Shangguan, Z.~Yang, Y.~Liu, Sleep hunter: Towards fine grained sleep
  stage tracking with smartphones, IEEE Transactions on Mobile Computing 15~(6)
  (2016) 1514--1527.
\newblock \href {https://doi.org/10.1109/TMC.2015.2462812}
  {\path{doi:10.1109/TMC.2015.2462812}}.

\bibitem{8718344}
Y.~Liu, Z.~Li, aleak: Context-free side-channel from your smart watch leaks
  your typing privacy, IEEE Transactions on Mobile Computing 19~(8) (2020)
  1775--1788.
\newblock \href {https://doi.org/10.1109/TMC.2019.2917659}
  {\path{doi:10.1109/TMC.2019.2917659}}.

\bibitem{8985418}
J.~Lin, J.~Niu, X.~Liu, M.~Guizani, Protecting your shopping preference with
  differential privacy, IEEE Transactions on Mobile Computing 20~(5) (2021)
  1965--1978.
\newblock \href {https://doi.org/10.1109/TMC.2020.2972001}
  {\path{doi:10.1109/TMC.2020.2972001}}.

\bibitem{8423084}
F.~Shen, J.~D. Vecchio, A.~Mohaisen, S.~Y. Ko, L.~Ziarek, Android malware
  detection using complex-flows, IEEE Transactions on Mobile Computing 18~(6)
  (2019) 1231--1245.
\newblock \href {https://doi.org/10.1109/TMC.2018.2861405}
  {\path{doi:10.1109/TMC.2018.2861405}}.

\bibitem{9000578}
S.~Chang, H.~Chen, H.~Zhu, X.~Hu, D.~Cao, Cosafe: Securing mobile devices
  through mutual mobility consistency verification, IEEE Transactions on Mobile
  Computing 20~(5) (2021) 1761--1772.
\newblock \href {https://doi.org/10.1109/TMC.2020.2974222}
  {\path{doi:10.1109/TMC.2020.2974222}}.

\bibitem{iOSPlay2020}
P.~Viswanathan,
  \href{https://www.lifewire.com/ios-app-store-vs-google-play-store-for-app-developers-2373130}{ios
  app store vs. google play store} (2020).
\newline\urlprefix\url{https://www.lifewire.com/ios-app-store-vs-google-play-store-for-app-developers-2373130}

\bibitem{CurryMarkets21}
D.~Curry.
\newblock \href{https://www.businessofapps.com/data/app-stores/}{[link]}.
\newline\urlprefix\url{https://www.businessofapps.com/data/app-stores/}

\bibitem{8595206}
H.~{Wang}, H.~{Li}, L.~{Li}, Y.~{Guo}, G.~{Xu}, Why are android apps removed
  from google play? a large-scale empirical study, in: 2018 IEEE/ACM 15th
  International Conference on Mining Software Repositories (MSR), 2018, pp.
  231--242.

\bibitem{PlayProtect2020}
Google, \href{https://www.android.com/play-protect/}{Google play protect}
  (2020).
\newline\urlprefix\url{https://www.android.com/play-protect/}

\bibitem{PlayStoreAgrement21}
G.~inc.,
  \href{https://play.google.com/about/developer-distribution-agreement.html}{Google
  play developer distribution agreement} (2021).
\newline\urlprefix\url{https://play.google.com/about/developer-distribution-agreement.html}

\bibitem{DeveloperPolicyCenter21}
Google,
  \href{https://play.google.com/about/developer-content-policy/}{Developer
  policy center} (2021).
\newline\urlprefix\url{https://play.google.com/about/developer-content-policy/}

\bibitem{6523820}
M.~{Kuehnhausen}, V.~S. {Frost}, Trusting smartphone apps? to install or not to
  install, that is the question, in: 2013 IEEE International Multi-Disciplinary
  Conference on Cognitive Methods in Situation Awareness and Decision Support
  (CogSIMA), 2013, pp. 30--37.
\newblock \href {https://doi.org/10.1109/CogSIMA.2013.6523820}
  {\path{doi:10.1109/CogSIMA.2013.6523820}}.

\bibitem{10.5555/2534766.2534812}
R.~Pandita, X.~Xiao, W.~Yang, W.~Enck, T.~Xie, Whyper: Towards automating risk
  assessment of mobile applications, in: Proceedings of the 22nd USENIX
  Conference on Security, SEC'13, USENIX Association, USA, 2013, p. 527–542.

\bibitem{7283021}
M.~{Gómez}, R.~{Rouvoy}, M.~{Monperrus}, L.~{Seinturier}, A recommender system
  of buggy app checkers for app store moderators, in: 2015 2nd ACM
  International Conference on Mobile Software Engineering and Systems, 2015,
  pp. 1--11.

\bibitem{DBLP:conf/icse/SlavinWHHKBBN16}
R.~Slavin, X.~Wang, M.~B. Hosseini, J.~Hester, R.~Krishnan, J.~Bhatia, T.~D.
  Breaux, J.~Niu, \href{https://doi.org/10.1145/2884781.2884855}{Toward a
  framework for detecting privacy policy violations in android application
  code}, in: Proceedings of the 38th International Conference on Software
  Engineering, {ICSE} 2016, Austin, TX, USA, May 14-22, {ACM}, 2016, pp.
  25--36.
\newblock \href {https://doi.org/10.1145/2884781.2884855}
  {\path{doi:10.1145/2884781.2884855}}.
\newline\urlprefix\url{https://doi.org/10.1145/2884781.2884855}

\bibitem{10.1145/2736277.2741084}
S.~Seneviratne, A.~Seneviratne, M.~A. Kaafar, A.~Mahanti, P.~Mohapatra,
  \href{https://doi.org/10.1145/2736277.2741084}{Early detection of spam mobile
  apps}, in: Proceedings of the 24th International Conference on World Wide
  Web, WWW '15, 2015, p. 949–959.
\newblock \href {https://doi.org/10.1145/2736277.2741084}
  {\path{doi:10.1145/2736277.2741084}}.
\newline\urlprefix\url{https://doi.org/10.1145/2736277.2741084}

\bibitem{chenatal16}
T.~Chen, C.~Guestrin,
  \href{http://dx.doi.org/10.1145/2939672.2939785}{Xgboost}, Proceedings of the
  22nd ACM SIGKDD International Conference on Knowledge Discovery and Data
  Mining (Aug 2016).
\newblock \href {https://doi.org/10.1145/2939672.2939785}
  {\path{doi:10.1145/2939672.2939785}}.
\newline\urlprefix\url{http://dx.doi.org/10.1145/2939672.2939785}

\bibitem{H0YJFT_2022}
F.~Mohsen, \href{https://doi.org/10.34894/H0YJFT}{{The manifest and store data
  of 870,515 Android mobile applications}} (2022).
\newblock \href {https://doi.org/10.34894/H0YJFT} {\path{doi:10.34894/H0YJFT}}.
\newline\urlprefix\url{https://doi.org/10.34894/H0YJFT}

\bibitem{androidobf}
S.~Dong, M.~Li, W.~Diao, X.~Liu, J.~Liu, Z.~Li, F.~Xu, K.~Chen, X.~Wang,
  K.~Zhang, \href{http://arxiv.org/abs/1801.01633}{Understanding android
  obfuscation techniques: {A} large-scale investigation in the wild}, CoRR
  abs/1801.01633 (2018).
\newblock \href {http://arxiv.org/abs/1801.01633} {\path{arXiv:1801.01633}}.
\newline\urlprefix\url{http://arxiv.org/abs/1801.01633}

\bibitem{PackageManager2021}
Google,
  \href{https://developer.android.com/reference/android/content/pm/PackageManager}{Android
  packagemanager} (2021).
\newline\urlprefix\url{https://developer.android.com/reference/android/content/pm/PackageManager}

\bibitem{PermissionsOnAndroid}
G.~inc.,
  \href{"https://developer.android.com/guide/topics/permissions/index.html"}{{Permissions
  on Android}} (April 2017).
\newline\urlprefix\url{"https://developer.android.com/guide/topics/permissions/index.html"}

\bibitem{10.1145/3442381.3449990}
F.~Lin, H.~Wang, L.~Wang, X.~Liu,
  \href{https://doi.org/10.1145/3442381.3449990}{A Longitudinal Study of
  Removed Apps in IOS App Store}, Association for Computing Machinery, New
  York, NY, USA, 2021, p. 1435–1446.
\newline\urlprefix\url{https://doi.org/10.1145/3442381.3449990}

\bibitem{10.1145/2873587.2873597}
M.~Liu, H.~Wang, Y.~Guo, J.~Hong,
  \href{https://doi.org/10.1145/2873587.2873597}{Identifying and analyzing the
  privacy of apps for kids}, in: Proceedings of the 17th International Workshop
  on Mobile Computing Systems and Applications, HotMobile '16, Association for
  Computing Machinery, New York, NY, USA, 2016, p. 105–110.
\newblock \href {https://doi.org/10.1145/2873587.2873597}
  {\path{doi:10.1145/2873587.2873597}}.
\newline\urlprefix\url{https://doi.org/10.1145/2873587.2873597}

\bibitem{10.1145/2295136.2295141}
B.~P. Sarma, N.~Li, C.~Gates, R.~Potharaju, C.~Nita-Rotaru, I.~Molloy,
  \href{https://doi.org/10.1145/2295136.2295141}{Android Permissions: A
  Perspective Combining Risks and Benefits}, ACM, New York, NY, USA, 2012, p.
  13–22.
\newline\urlprefix\url{https://doi.org/10.1145/2295136.2295141}

\bibitem{8455897}
S.~M. {Habib}, N.~{Alexopoulos}, M.~M. {Islam}, J.~{Heider}, S.~{Marsh},
  M.~{Müehlhäuser}, Trust4app: Automating trustworthiness assessment of
  mobile applications, in: 2018 17th IEEE International Conference On Trust,
  Security And Privacy In Computing And Communications/ 12th IEEE International
  Conference On Big Data Science And Engineering (TrustCom/BigDataSE), 2018,
  pp. 124--135.
\newblock \href {https://doi.org/10.1109/TrustCom/BigDataSE.2018.00029}
  {\path{doi:10.1109/TrustCom/BigDataSE.2018.00029}}.

\bibitem{Sarma:2012:APP:2295136.2295141}
B.~P. Sarma, N.~Li, C.~Gates, R.~Potharaju, C.~Nita-Rotaru, I.~Molloy,
  \href{http://doi.acm.org/10.1145/2295136.2295141}{Android permissions: A
  perspective combining risks and benefits}, in: Proceedings of the 17th ACM
  Symposium on Access Control Models and Technologies, SACMAT '12, ACM, New
  York, NY, USA, 2012, pp. 13--22.
\newblock \href {https://doi.org/10.1145/2295136.2295141}
  {\path{doi:10.1145/2295136.2295141}}.
\newline\urlprefix\url{http://doi.acm.org/10.1145/2295136.2295141}

\bibitem{8456016}
F.~Mohsen, H.~Abdelhaq, H.~Bisgin, A.~Jolly, M.~Szczepanski, Countering
  intrusiveness using new security-centric ranking algorithm built on top of
  elasticsearch, in: 2018 17th IEEE International Conference On Trust, Security
  And Privacy In Computing And Communications/ 12th IEEE International
  Conference On Big Data Science And Engineering (TrustCom/BigDataSE), 2018,
  pp. 1048--1057.
\newblock \href {https://doi.org/10.1109/TrustCom/BigDataSE.2018.00147}
  {\path{doi:10.1109/TrustCom/BigDataSE.2018.00147}}.

\bibitem{10.1145/3308558.3313611}
H.~Wang, H.~Li, Y.~Guo,
  \href{https://doi-org.proxy-ub.rug.nl/10.1145/3308558.3313611}{Understanding
  the evolution of mobile app ecosystems: A longitudinal measurement study of
  google play}, in: WWW, WWW '19, ACM, New York, NY, USA, 2019, p. 1988–1999.
\newblock \href {https://doi.org/10.1145/3308558.3313611}
  {\path{doi:10.1145/3308558.3313611}}.
\newline\urlprefix\url{https://doi-org.proxy-ub.rug.nl/10.1145/3308558.3313611}

\bibitem{6735264}
N.~{Peiravian}, X.~{Zhu}, Machine learning for android malware detection using
  permission and api calls, in: 2013 IEEE 25th International Conference on
  Tools with Artificial Intelligence, 2013, pp. 300--305.

\bibitem{6298136}
D.~{Wu}, C.~{Mao}, T.~{Wei}, H.~{Lee}, K.~{Wu}, Droidmat: Android malware
  detection through manifest and api calls tracing, in: 2012 Seventh Asia Joint
  Conference on Information Security, 2012, pp. 62--69.

\bibitem{doi:10.1080/01969722.2013.803889}
B.~Sanz, I.~Santos, C.~Laorden, X.~Ugarte-Pedrero, J.~Nieves, P.~G. Bringas,
  G.~Álvarez Marañón,
  \href{https://doi.org/10.1080/01969722.2013.803889}{Mama: Manifest analysis
  for malware detection in android}, Cybernetics and Systems 44~(6-7) (2013)
  469--488.
\newblock \href
  {http://arxiv.org/abs/https://doi.org/10.1080/01969722.2013.803889}
  {\path{arXiv:https://doi.org/10.1080/01969722.2013.803889}}, \href
  {https://doi.org/10.1080/01969722.2013.803889}
  {\path{doi:10.1080/01969722.2013.803889}}.
\newline\urlprefix\url{https://doi.org/10.1080/01969722.2013.803889}

\bibitem{Sato2013DetectingAM}
R.~Sato, D.~Chiba, S.~Goto, Detecting android malware by analyzing manifest
  files, in: Proceedings of the Asia-Pacific Advanced Network, Vol.~36, 2013,
  p.~23.
\newblock \href {https://doi.org/10.7125/APAN.36.4}
  {\path{doi:10.7125/APAN.36.4}}.

\bibitem{7035780}
S.~{Feldman}, D.~{Stadther}, B.~{Wang}, Manilyzer: Automated android malware
  detection through manifest analysis, in: 2014 IEEE 11th International
  Conference on Mobile Ad Hoc and Sensor Systems, 2014, pp. 767--772.

\bibitem{7790261}
X.~{Li}, J.~{Liu}, Y.~{Huo}, R.~{Zhang}, Y.~{Yao}, An android malware detection
  method based on androidmanifest file, in: 2016 4th International Conference
  on Cloud Computing and Intelligence Systems (CCIS), 2016, pp. 239--243.

\bibitem{10.1145/3038912.3052712}
H.~Wang, Z.~Liu, Y.~Guo, X.~Chen, M.~Zhang, G.~Xu, J.~Hong,
  \href{https://doi.org/10.1145/3038912.3052712}{An explorative study of the
  mobile app ecosystem from app developers' perspective}, in: Proceedings of
  the 26th International Conference on World Wide Web, WWW '17, International
  World Wide Web Conferences Steering Committee, Republic and Canton of Geneva,
  CHE, 2017, p. 163–172.
\newblock \href {https://doi.org/10.1145/3038912.3052712}
  {\path{doi:10.1145/3038912.3052712}}.
\newline\urlprefix\url{https://doi.org/10.1145/3038912.3052712}

\bibitem{Alsahaf2019}
A.~Alsahaf, G.~Azzopardi, B.~Ducro, E.~Hanenberg, R.~F. Veerkamp, N.~Petkov,
  Estimation of muscle scores of live pigs using a kinect camera, IEEE Access 7
  (2019) 52238--52245.
\newblock \href {https://doi.org/10.1109/ACCESS.2019.2910986}
  {\path{doi:10.1109/ACCESS.2019.2910986}}.

\bibitem{FARRUGIA2020113318}
S.~Farrugia, J.~Ellul, G.~Azzopardi,
  \href{https://www.sciencedirect.com/science/article/pii/S0957417420301433}{Detection
  of illicit accounts over the ethereum blockchain}, Expert Systems with
  Applications 150 (2020) 113318.
\newblock \href {https://doi.org/https://doi.org/10.1016/j.eswa.2020.113318}
  {\path{doi:https://doi.org/10.1016/j.eswa.2020.113318}}.
\newline\urlprefix\url{https://www.sciencedirect.com/science/article/pii/S0957417420301433}

\bibitem{Lovdal2020}
S.~S. Lövdal, R.~J.~D. Hartigh, G.~Azzopardi,
  \href{http://journals.humankinetics.com/view/journals/ijspp/aop/article-10.1123-ijspp.2020-0518/article-10.1123-ijspp.2020-0518.xml}{Injury
  prediction in competitive runners with machine learning}, International
  Journal of Sports Physiology and Performance (01 Jan. 2020) 1 -- 10\href
  {https://doi.org/10.1123/ijspp.2020-0518}
  {\path{doi:10.1123/ijspp.2020-0518}}.
\newline\urlprefix\url{http://journals.humankinetics.com/view/journals/ijspp/aop/article-10.1123-ijspp.2020-0518/article-10.1123-ijspp.2020-0518.xml}

\bibitem{Rice2005}
M.~Rice, G.~Harris, {Comparing effect sizes in follow-up studies: ROC area,
  Cohen's d, and r}, {Law and Human Behaviour} {29}~({5}) ({2005}) {615--620}.
\newblock \href {https://doi.org/{10.1007/s10979-005-6832-7}}
  {\path{doi:{10.1007/s10979-005-6832-7}}}.

\bibitem{10.1145/3278532.3278558}
H.~Wang, Z.~Liu, J.~Liang, N.~Vallina-Rodriguez, Y.~Guo, L.~Li, J.~Tapiador,
  J.~Cao, G.~Xu,
  \href{https://doi-org.proxy-ub.rug.nl/10.1145/3278532.3278558}{Beyond google
  play: A large-scale comparative study of chinese android app markets}, in:
  Proceedings of the Internet Measurement Conference 2018, IMC '18, Association
  for Computing Machinery, New York, NY, USA, 2018, p. 293–307.
\newblock \href {https://doi.org/10.1145/3278532.3278558}
  {\path{doi:10.1145/3278532.3278558}}.
\newline\urlprefix\url{https://doi-org.proxy-ub.rug.nl/10.1145/3278532.3278558}

\end{thebibliography}

%\pagebreak

\end{document}